\documentclass[12pt]{article}
\usepackage{amsmath,amssymb,amsthm}
\usepackage{latexsym,graphicx,color,subfigure,mathrsfs}
\usepackage[x11names]{xcolor}
\usepackage[colorlinks,linkcolor=purple,citecolor=teal]{hyperref}
\usepackage[compress,square,numbers]{natbib}

\newcommand{\beq}{\begin{eqnarray}}
\newcommand{\eeq}{\end{eqnarray}}
\newcommand{\non}{\nonumber\\}
\newcommand{\U}{\mbox{U}}
\newcommand{\SU}{\mbox{SU}}

\newcommand{\SO}{\mbox{SO}}
\newcommand{\bm}{\mathbf{m}}
\newcommand{\br}{\mathbf{r}}

\newcommand{\by}{\mathbf{y}}
\newcommand{\bz}{\mathbf{z}}

\newcommand{\bY}{\mathbf{Y}}

\newcommand{\bsigma}{\boldsymbol{\sigma}}
\newcommand{\bOmega}{\boldsymbol{\Omega}}
\newcommand{\bell}{\boldsymbol{\ell}}
\def\calE{{\cal E}}

\def\calM{{\cal M}}
\def\calN{{\cal N}}
\def\calL{{\cal L}}
\def\Tr{\mathrm{Tr}}
\def\sign{\mathop{\mathrm{sign}}\nolimits}
\def\p{\partial}
\renewcommand{\d}{{\mathrm{d}}}
\renewcommand{\i}{\mathrm{i}}

\newtheorem{prop}{Proposition}

\newtheorem{lemma}{Lemma}
\newtheorem{theorem}{Theorem}
\newtheorem{corollary}{Corollary}

\makeatletter
\@addtoreset{equation}{section}

\renewcommand{\thefootnote}{\fnsymbol{footnote}}
\makeatother

\makeatletter
\newcommand{\thetablename}{Table}
\def\fnum@table{\thetablename\ \thetable}
\makeatother

\begin{document}
\pagenumbering{Roman}
\thispagestyle{empty}
\begin{flushright}
YGHP-20-06
\end{flushright}
\vspace{3mm}

\begin{center}
{\Large \bf 1/2-BPS vortex strings in $\mathcal{N}=2$ supersymmetric\\ $\U(1)^N$ gauge theories} 
\\[15mm]
{Sven Bjarke~{\sc Gudnason}}$^{1}$\footnote{\it Corresponding author; e-mail address:
gudnason(at)henu.edu.cn},
{Minoru~{\sc Eto}}$^{2}$,
{Muneto~{\sc Nitta}}$^{3}$
\vskip 6 mm

\bigskip\bigskip
{\it
${}^1$Institute of Contemporary Mathematics, School of
  Mathematics and Statistics,\\ Henan University, Kaifeng, Henan 475004,
  P.~R.~China\\
${}^2$Department of Physics, Yamagata University,
  Kojirakawa-machi 1-4-12, Yamagata,\\ Yamagata 990-8560, Japan\\
${}^3$Department of Physics, and Research and Education Center
  for Natural Sciences,\\ Keio University, Hiyoshi 4-1-1, Yokohama,
  Kanagawa 223-8521, Japan
}

\bigskip

\bigskip

{\bf Abstract}\\[5mm]
{\parbox{13cm}{\hspace{5mm}
Strings in $\mathcal{N}=2$ supersymmetric $\U(1)^N$ gauge
theories with $N$ hypermultiplets are studied in the generic setting
of an arbitrary Fayet-Iliopoulos triplet of parameters for each gauge 
group and an invertible charge matrix.
Although the string tension is generically of a square-root form, it
turns out that all existing BPS (Bogomol'nyi-Prasad-Sommerfield)
solutions have a tension which is 
linear in the magnetic fluxes, which in turn are linearly related to
the winding numbers.
The main result is a series of theorems establishing three different
kinds of solutions of the so-called constraint equations, which can be
pictured as orthogonal directions to the magnetic flux in $\SU(2)_R$
space.
We further prove for all cases, that a seemingly vanishing
Bogomol'nyi bound cannot have solutions.
Finally, we write down the most general vortex equations in both
master form and Taubes-like form.
Remarkably, the final vortex equations essentially look
Abelian in the sense that there is no trace of the $\SU(2)_R$
symmetry in the equations, after the constraint equations have been
solved. 
}}
\end{center}
\newpage
\pagenumbering{arabic}
\setcounter{page}{1}
\setcounter{footnote}{0}
\renewcommand{\thefootnote}{\arabic{footnote}}


\tableofcontents

\section{Introduction and summary}

In string theory there exists two kinds of string, namely fundamental
strings or F-strings and D1-branes or D-strings
\cite{Polchinski:1995mt,Witten:1995im,Copeland:2003bj,Polchinski:2004ia,Polchinsky:1998}. 
Bound states between those two kinds of string yield a very
distinctive tension formula
\beq
T \propto \sqrt{p^2 + q^2/g_s^2},
\label{eq:string_theory_tension}
\eeq
where $p$ is the number of F-strings and $q$ the number of D-strings
and $g_s$ is the string coupling constant.
A simple gauge theory mimicking $(p,q)$-string bound states was
proposed by Saffin \cite{Saffin:2005cs}, but the string tension is
unfortunately linear in the two winding numbers:
\beq
T \propto p + \mu q,
\label{eq:linear_tension}
\eeq
with $\mu>0$ a positive constant.
Jackson then found that the square-root tension formula actually
exists naturally in $\mathcal{N}=2$ supersymmetric gauge theories
\cite{Jackson:2006qc}, where the strings are bound states of D-term
strings and F-term strings and the tension is exactly like
Eq.~\eqref{eq:string_theory_tension}.
This happy discovery came to an early halt, as Jackson showed in the
same paper that no BPS\footnote{BPS is an abbreviation for
  Bogomol'nyi-Prasad-Sommerfield who where the authors first discovering a
  reduction from second-order Euler-Lagrange equations for a magnetic
  monopole to first-order partial differential equations (PDEs), now
  known as BPS equations. BPS equations and solutions are now used in
  many areas of theoretical physics, from gauge theory to string
  theory. } solutions -- 
with the square-root tension -- seemed to exist.

The gauge theory that Jackson studied is a $\U(1)^N$ $\mathcal{N}=2$
supersymmetric gauge theory with $N$ hypermultiplets. Not all the
field content will be used in the construction of vortices, so
therefore it is simply $N$ $\U(1)$ gauge fields with $N$ pairs of
complex scalar fields.
It holds true that if a flavor of scalar fields possesses winding
(topological charge), then either of the two (complex) scalar fields
in the pair must vanish everywhere.
This can readily be realized by analyzing the self-dual equations,
which are equal for the two scalars in the pair, except for having
opposite (electric) charges.
We confirm this fact, but do not restrict to the case that all winding
numbers must be nonzero, which relaxes this statement. 
In Jackson's analysis, it was found that the magnetic fluxes should be
proportional to the column-vectors of the inverse of the charge
matrix.
Unfortunately, this is impossible, as if true, then the determinant of
the inverse of the charge matrix vanishes and hence the charge matrix
does not exist!
Another option was pointed out, however, which is to align all the
vectors of Fayet-Iliopoulos (FI) parameters in $\SU(2)_R$ space (which
roughly is the space containing the D-term in one direction and the
F-term in further two directions).
This option, however, does not possess the square-root tension, but
the normal linear tension formula \eqref{eq:linear_tension}.
We confirm this possibility as a solution type and call them solutions
of type A.
The final step in Jackson's analysis concludes from analyzing the
asymptotic behavior of the fields that the charge matrix must be
diagonal in order to have BPS solutions.
We do not agree with this conclusion and we are able to find
BPS-saturated solutions with non-diagonal charge matrices.

We consider briefly the vacuum solutions (vacuum expectation values
(VEVs)) of the more general theory with $M$ hypermultiplets, but
conclude that we either have vacuum moduli (for $M>N$) or unbroken
gauge symmetry (for $M<N$), unless $M=N$\footnote{Actually, there is a
vacuum modulus (a complex phase) for each gauge group in the case of
$M=N$, which corresponds to the base point of the broken $\U(1)$
symmetry. This parameter has no physical relevance. }.
Hence this is the case we will consider in this paper. It also makes
the charge matrix a square matrix, which when possessing a
nonvanishing determinant, is invertible.
We will make this a requirement throughout the paper.
First, we will perform an $\SU(2)_R$ rotation of the fields to put the
FI vectors on a standard form, so that as many parameters as possible
vanish -- without loss of generality.
Then we will consider the most general supersymmetry projection for a
string pointed in the spatial $x^3$ direction and this gives the BPS
equations that we will study.
We confirm these BPS equations by writing down the Bogomol'nyi
completion, which in turn yields the Bogomol'nyi bound, which is
saturated for BPS solutions.
Curiously, it is not clear that the Bogomol'nyi bound is always
nonvanishing.
We will, however, prove for all types of solutions, that the vanishing
Bogomol'nyi bound does not possess any BPS solutions.
When working with the BPS equations, it turns out to be convenient to
perform a change of basis to a diagonal basis, which we shall call the
$\Psi$ basis, as opposed to the canonical basis, which we call the
$\Phi$ basis.
In the $\Psi$ basis, we confirm Jackson's result that either of the
two fields in a (hypermultiplet) pair must vanish when the winding
number is nonzero, see lemma \ref{lemma:1}.
The BPS equations are not one equation for the magnetic flux equating
the square-root of the potential as usual in $\mathcal{N}=1$ gauge
theories, but contain another two equations in the $\Psi$ basis, which
we shall coin constraint equations.
A main result is to split the solutions of the constraint equation
into two different types, which we shall denote type A and type B, see
proposition \ref{prop:1}.
In type A solutions, all the FI vectors are parallel (see theorem
\ref{thm:1}), but in type B at least one vector is not parallel with
the others.
A further constraint comes from the fact that there exists no
holomorphic $\U(1)$ bundles of negative degree and hence if the VEV is
positive, the ``fundamental field'' must carry the winding, but if it
is negative, the ``anti-fundamental fields'' must carry the winding.
This is summarized in theorem \ref{thm:2} for type A solutions.
With this result in hand, we can now prove for type A solutions that
no solutions exist in the case where the Bogomol'nyi bound vanishes,
see theorem \ref{thm:3}.

For type B solutions the FI vectors are not all parallel.
This means that the product of some pair of fields does not vanish in
the vacuum and hence by the above holomorphicity constraint (coming
from the pair of self-dual equations), those pairs cannot possess
winding -- we shall dub them non-winding fields or inert fields. 
We find that there are two different solutions to the constraint
equations in the type B case, see theorem \ref{thm:4} and one is given
by allowing only a single field to wind. This will be the type B1
solution. 
The other option is to restrict the charge matrix to be of a
block-diagonal form between the winding and the non-winding fields. 
Exactly this form will avoid the mixing of fields observed by Jackson.
For the winding fields of type B2, everything is analogous to the
solutions of type A, but we prove every step carefully.
First, we find that the FI vectors coupled to the winding fields,
must be parallel, see lemma \ref{lemma:2}. 
We further prove that the non-winding fields in type B2 satisfy both
the constraint equations (lemma \ref{lemma:3}) and would-be vortex
equations (theorem \ref{thm:5}) by means of their vacuum solution.
This is possible only because of the restricted form of the charge
matrix. 
We find, without loss of generality, that we can rotate the basis and
relabel the gauge groups to obtain parallel FI vectors with real
vacuum solutions, see theorem \ref{thm:6} and corollary \ref{coro:5}.
At this point, the same argument as in the type A case can be used to
relate the sign of the vacuum solution to the sign of the winding
number, see theorem \ref{thm:7}.
In turn we can prove analogously that the vanishing Bogomol'nyi bound
has no solutions, see theorem \ref{thm:8}.
For the solutions of type B1, we still have to set either of the
fields in the hypermultiplet pair to zero by corollary \ref{coro:4},
but otherwise have no further constraints.
Because of the simplicity of having only a single winding field, we
are able to prove also in this last case, that the vanishing
Bogomol'nyi bound has no solutions, see theorem \ref{thm:9}.

Finally, we are in a position to write down the governing equations,
which we call the master equations and they are given for type A, type
B2 and type B1 solutions in theorems \ref{thm:10}, \ref{thm:11} and
\ref{thm:12}, respectively.
Since this form of the equations may be unfamiliar to many readers, we
provide a change of variables to a form more similar to the Taubes
equation in Sec.~\ref{sec:Taubes_eqs}.
We give some explicit examples in Sec.~\ref{sec:examples}.
It is remarkable, that the vortex equations of type A and type B2 are
identical to the vortex equations found in Ref.~\cite{Schroers:1996zy}
in the context of essentially the same model as studied here, but with
only $\mathcal{N}=1$ supersymmetry and hence no $\SU(2)_R$ structure.
We may say that after solving the constraint equations in the
$\mathcal{N}=2$ theory, the vortex equations look basically like those
of an $\mathcal{N}=1$ theory.

As for the string tension, it turns out always to be of the linear
form \eqref{eq:linear_tension} and not of the square-root form
\eqref{eq:string_theory_tension}.
However, we find nontrivial BPS solutions which can have arbitrary
charge matrices, but alignment of the FI vectors is a necessity for
all winding flavors (in case there is more than one winding flavor). 

Physically, we may interpret the vanishing Bogomol'nyi bound as a
signal that the theory touches the Coulomb branch and hence possesses
unbroken gauge symmetry, which we do not allow. 

Apart from the situation with $M>N$, where we would have vacuum moduli
and semi-local types of string \cite{Vachaspati:1991dz}, an
interesting possibility has been left out in this work.
That is, the case of the charge matrix with a vanishing determinant.
It completely invalidates all analysis done in this paper, so we
cannot say anything about such cases on the basis of our analysis.
However, it is not necessarily unphysical to consider systems with
charge matrices that do not possess an inverse (i.e.~with vanishing
determinant).
The simplest two cases would be if all flavors have the same charge
under each gauge group and the other would be that one flavor has the
same charge under all gauge groups (but not necessarily the same as
other flavors).
We leave this case for future work.

\section{The model}

We consider an $\calN=2$ supersymmetric $\U(1)^N$ gauge 
theory in $3+1$ dimensions with $M$ ``flavors'' (hypermultiplets)
$\Phi_A=(\phi_A,\tilde\phi_A^*)^{\rm T}$, $A=1,2,\ldots,M$, with the
following bosonic sector
\begin{align}
  \calL &=
  -\sum_a\frac{1}{4e_a^2}F_{\mu\nu}^a F^{a\mu\nu}
  -\sum_A D_\mu\Phi_A^\dag D^\mu\Phi_A
  -\sum_a\frac{1}{e_a^2}\p_\mu\Sigma^a\p^\mu\Sigma^a \non
  &\phantom{=\ }
  -\sum_{a,\alpha}\frac{1}{2e_a^2}\left(Y_a^\alpha\right)^2
  -\sum_A\Big(\sum_a Q_{Aa}\Sigma^a\Big)^2\Phi_A^\dag\Phi_A.
  \label{eq:L}
\end{align}
The neutral scalars $\Sigma^a$ belong to the $\calN=2$ vector
multiplet together with the photons $A_\mu^a$ which have the field
strength
\beq
F_{\mu\nu}^a = \p_\mu A_\nu^a - \p_\nu A_\mu^a,
\eeq
the gauge couplings are $e_a>0$, and finally the index $a=1,2,\ldots,N$
corresponds to the gauge groups $\U(1)_a$.
The gauge covariant derivative for the complex scalar fields is
\beq
D_\mu\Phi_A = \p_\mu\Phi_A + \i\sum_a Q_{Aa} A_\mu^a \Phi_A,
\eeq
where $Q_{Aa}$ is a charge matrix with flavor and gauge index.
Finally,
\beq
Y_a^\alpha \equiv
e_a^2\bigg(\sum_A \Phi_A^\dag\sigma^\alpha\Phi_A Q_{Aa} - r_a^\alpha\bigg),
\label{eq:Y}
\eeq
is an $\SU(2)_R$ vector of the D-terms ($\alpha=3$) and the F-terms
($\alpha=1,2$), where $\sigma^\alpha$ are the standard Pauli matrices
and $r_a^\alpha$ are $N$ Fayet-Iliopoulos (FI) triplets of
parameters. 
Often we will write $\SU(2)_R$ vectors as a boldfaced vector (e.g.~$\br_a$)
(suppressing the $\SU(2)_R$ index $\alpha$) and use standard vector
notations, like the dot product.
The indices $\mu,\nu$ are spacetime indices and are summed over with
the Einstein summation convention (that is, any couple of indices are
automatically summed over $\mu=0,1,2,3$).
The internal indices, i.e.~the gauge group index $a=1,2,\ldots,N$, the
flavor index $A=1,2,\ldots,M$ and the $\SU(2)_R$ index $\alpha=1,2,3$
are \emph{not} summed over by the Einstein convention throughout the
paper and if a sum is intended, we write the sum explicitly. 
The metric signature used here is the mostly positive one, convenient
for solitons.

The real fields $\Sigma^a$ are source-free fields and are not used in
the construction of vortices. The minimum of the energy is attained by
the vacuum solution for $\Sigma^a$ which is $\Sigma^a=0$.
Hence in the remainder of the paper, we will set $\Sigma^a=0$. This
yields the reduced Lagrangian density 
\begin{align}
  \calL &=
  -\sum_a\frac{1}{4e_a^2}F_{\mu\nu}^a F^{a\mu\nu}
  -\sum_A D_\mu\Phi_A^\dag D^\mu\Phi_A
  -\sum_{a,\alpha}\frac{e_a^2}{2}\Big(\sum_A
  \Phi_A^\dag\sigma^\alpha\Phi_A Q_{Aa} - r_a^\alpha\Big)^2.
  \label{eq:Lreduced}
\end{align}
We assume that each $\U(1)$ gauge group is compact and thus all
elements of the charge matrix $Q_{Aa}\in\mathbb{Z}$ are integers.

Each $\U(1)_a$ gauge group corresponds a redundancy known as a gauge
symmetry. The symmetry acts on the fields as the following
transformation
\begin{align}
  \Phi_A &\to \Phi_A \exp\bigg(\i\sum_a Q_{Aa} \alpha_a(x)\bigg), \qquad
  \alpha_a(x)\in\mathbb{R},\non
  A_\mu^a &\to A_\mu^a - \p_\mu \alpha_a(x),
\end{align}
which leaves the Lagrangian \eqref{eq:L} (and \eqref{eq:Lreduced})
invariant.

\subsection{The FI parameters}

We will now perform an $\SU(2)_R$ transformation on the scalar fields,
$\Phi\to U\Phi$, with $U$ a constant unitary matrix. This leaves the
kinetic term invariant.
The potential transforms, however, and in particular we get
\beq
\sum_A \Phi_A^\dag\sigma^\alpha\Phi_A Q_{Aa} - r_a^\alpha
\quad\longrightarrow\quad
\sum_A \Phi_A^\dag U^\dag\sigma^\alpha U\Phi_A Q_{Aa} - r_a^\alpha,
\eeq
and using that $U^\dag\sigma^\alpha U$ takes value in the
$\mathfrak{su}(2)_R$ algebra, we can write
\beq
U^\dag\sigma^\alpha U
= \frac12\sum_\beta\Tr[U^\dag\sigma^\alpha U\sigma^\beta]\sigma^\beta
\equiv \sum_\beta R_{\alpha\beta} \sigma^\beta.
\eeq
Using that $R\in\SO(3)$ and in particular
$\sum_{\alpha}R_{\alpha\beta}R_{\alpha\gamma}=\delta_{\beta\gamma}$,
we can write 
\begin{align}
\sum_{a,\alpha}\frac{e_a^2}{2}\Big(
\sum_A \Phi_A^\dag U^\dag\sigma^\alpha U\Phi_A Q_{Aa} - r_a^\alpha
\Big)^2
&= \sum_{a,\alpha}\frac{e_a^2}{2}\bigg(
\sum_\beta R_{\alpha\beta}\Big(\sum_A \Phi_A^\dag\sigma^\beta\Phi_A Q_{Aa} - \sum_\gamma r_a^\gamma R_{\gamma\beta}\Big)
\bigg)^2 \non
&= \sum_{a,\alpha}\frac{e_a^2}{2}\Big(
\sum_A \Phi_A^\dag\sigma^\alpha\Phi_A Q_{Aa} - \tilde{r}_a^\alpha
\Big)^2,
\end{align}
with the rotated FI parameters
\beq
\tilde{r}_a^\alpha\equiv \sum_{\beta}r_a^\beta R_{\beta\alpha}.
\eeq
Now we can simplify the FI parameters using the $\SU(2)_R$ rotations,
depending on the total number of gauge groups, $N$.
We assume that for each gauge group, there is an FI vector with
nonvanishing length
\beq
\br_a\cdot\br_a > 0, \qquad \forall a=1,2,\ldots,N,
\label{eq:FInonzerolength}
\eeq
where $\br_a$ is a 3-vector in $\SU(2)_R$ space for each gauge group
$\U(1)_a$.

The first vector, say $\br_1$, can be rotated to
\beq
\tilde{\br}_1 = (\alpha,0,0), \qquad \alpha>0,
\label{eq:r1}
\eeq
and using the residual $\U(1)\subset\SU(2)_R$ symmetry, we can rotate
the next vector to
\beq
\tilde{\br}_2 = (\beta,\gamma,0), \qquad \beta\in\mathbb{R}, \quad\gamma\geq 0.
\label{eq:r2}
\eeq
These two transformations use up our freedom to rotate within
$\SU(2)_R$ and the remaining FI parameters (vectors) point in
arbitrary directions.
We will henceforth drop the tildes on the FI parameters.

The above considerations show that if we only consider two gauge
groups ($N=2$), then without loss of generality, the FI parameters
can be chosen to be in the $(\sigma^1,\sigma^2)$-plane.
For $N>2$, however, we need the full 3-dimensional $\SU(2)_R$ space
for the FI parameters.

\subsection{The vacuum and the charge matrix}

The vacuum equations are
\beq
\sum_A \langle\Phi_A\rangle^\dag\sigma^\alpha\langle\Phi_A\rangle
Q_{Aa} = r_a^\alpha, \qquad
\forall\alpha=1,2,3,\quad \forall a=1,2,\ldots,N,
\eeq
where $\langle\Phi_A\rangle$ denotes the vacuum expectation value
(VEV) of the complex scalar field doublet of flavor $A$. 
Let us first consider the number of variables versus the number of
constraints in this equation.
Each flavor of scalar field doublets has 4 real components yielding a
total of $4M$ degrees of freedom in the vacuum equation.
The number of constraints on the other hand are 3 for each gauge
group, so a total of $3N$ constraints.
For each gauge group, there is 1 vacuum modulus -- a free parameter. 
Thus if $N>M$ then generically the vacuum equations are
over-determined in which case one should minimize the potential (as it
cannot vanish unless enough of the FI parameters are equal to each
other). 
This, however, will lift the vacuum energy and break supersymmetry, so
we will not consider such case further in this paper.
On the other hand, if $N<M$ there will be further vacuum moduli. These
are typical characteristics of supersymmetric theories. 

We will require the gauge symmetry to be completely broken which
corresponds to the mass term for the photons
\beq
\frac12\sum_{a,b}(\mathcal{M}^2)_{ab} A_\mu^a A^{\mu b}, \qquad
(\mathcal{M}^2)_{ab} \equiv 2\sum_A e_a e_b 
  \langle\Phi_A\rangle^\dag\langle\Phi_A\rangle Q_{Aa} Q_{Ab},
\eeq
not having one or more zero eigenvalues:
\beq
\det (\mathcal{M}^2) > 0,
\eeq
where $e_a$ are positive definite gauge coupling constants.

We will now consider the solutions to the vacuum equations, as they
will be useful later.
Defining
\beq
\langle\Phi_A\rangle \equiv
\begin{pmatrix}
  \rho_A e^{-\i\vartheta_A}\\
  \tilde\rho_A e^{\i\tilde\vartheta_A}
\end{pmatrix},
\label{eq:Phi_vac}
\eeq
we get for each gauge group, $a=1,\ldots,N$, the vacuum equations
\begin{align}
2\sum_A \rho_A\tilde\rho_A
\exp\big[\i(\vartheta_A+\tilde\vartheta_A)\big] Q_{Aa} &= r_a^1+\i r_a^2
\equiv r_a,\label{eq:vac12}\\
\sum_A (\rho_A^2 - \tilde\rho_A^2) Q_{Aa} &= r_a^3.\label{eq:vac3}
\end{align}

The mass-squared matrix for the photons in terms of the new vacuum
variables reads
\beq
(\mathcal{M}^2)_{ab} = 2\sum_A e_a e_b 
\left(\rho_A^2 + \tilde\rho_A^2\right) Q_{Aa} Q_{Ab}.
\eeq
The computation of the determinant of this matrix depends on $N$ and
$M$.

In the case of $N=M$, it is easy to show that the determinant of the
mass-squared matrix reads
\beq
\det(\mathcal{M}^2) = 2^N\bigg(\prod_a e_a^2\bigg)
(\det Q)^2\prod_A(\rho_A^2+\tilde\rho_A^2).
\label{eq:detMsqN=M}
\eeq
In order to Higgs the gauge symmetry completely, we must have
$\det Q\neq 0$, all couplings positive $e_a>0$ as well as each flavor
must have a nonvanishing VEV in either the fundamental or the
antifundamental fields (or in both). 

If $N>M$ we have
\begin{equation}
\det(\mathcal{M}^2) =
2^N\!\sum_{b1,\cdots,b_N}\!\epsilon_{b_1\cdots b_N}
\left[e_1 e_{b_1}\sum_{A_1} Q_{A_11}Q_{A_1b_1}
  R_{A_1}^2\right]
\cdots
\left[e_N e_{b_N}\sum_{A_N} Q_{A_NN}Q_{A_Nb_N}
R_{A_N}^2\right],
\end{equation}
where we have defined $R_{A}^2\equiv\rho_{A}^2+\tilde\rho_{A}^2$.
In order for the determinant not to vanish, we must have a different
flavor for each bracket, as the same flavor is eliminated by the
epsilon tensor. For $N>M$, that is impossible and the determinant
always vanishes. For $N=M$, we get the result of
Eq.~\eqref{eq:detMsqN=M}.

For $N<M$, the determinant can be nonvanishing. It will be instructive
to consider first the case of $M=N+1$. The result will be a sum of
$M$ minor determinants
\beq
\det(\mathcal{M}^2) =
2^N\bigg(\prod_a e_a^2\bigg)
\sum_{B=1}^M (\det Q^{B})^2 \prod_{A\neq B}^M
(\rho_{A}^2 + \tilde\rho_{A}^2),
\eeq
where $\det Q^B$ is the minor determinant where the $B$th row is
removed.
The general case of $N<M$ thus reads
\begin{equation}
\det(\mathcal{M}^2) =
2^N\bigg(\prod_a e_a^2\bigg)
\sum_{B_1=1}^M \sum_{B_2>B_1}^M \cdots \sum_{B_{M-N}>B_{M-N-1}}^M 
(\det Q^{B_1\cdots B_{M-N}})^2
\prod_{A\neq B_1,B_2,\cdots,B_{M-N}}^M
(\rho_{A}^2 + \tilde\rho_{A}^2),
\end{equation}
where $\det Q^{B_1\cdots B_{M-N}}$ is the minor determinant with the
rows $B_1,\ldots,B_{M-N}$ removed. 
We shall not consider this case further in this paper. 

From this point on, we shall consider only the case $N=M$.
It will be convenient in the remainder of the paper to define two new
variables for solving the vacuum equations
\beq
z_A \equiv 2\rho_A\tilde\rho_A e^{\i(\vartheta_A+\tilde\vartheta_A)},\qquad
y_A \equiv \rho_A^2 - \tilde\rho_A^2,
\label{eq:zA_yA_def}
\eeq
in terms of which the general vacuum solution reads
\beq
z_A = \sum_a (r_a^1 + \i r_a^2) Q^{-1}_{aA}, \qquad
y_A = \sum_a r_a^3 Q^{-1}_{aA},
\label{eq:general_vac_sol}
\eeq
where it is understood that $Q^{-1}_{aA}$ is the inverse matrix of
$Q_{Aa}$. 
The following relations will come in handy in various computations
\beq
\langle\phi_A\tilde\phi_A\rangle = \frac12z_A^*, \qquad
|\langle\phi_A\rangle|^2 - |\langle\tilde\phi_A\rangle|^2 = y_A, \qquad
|\langle\phi_A\rangle|^2 + |\langle\tilde\phi_A\rangle|^2 = \sqrt{y_A^2 + |z_A|^2}.
\label{eq:useful_vac_rels}
\eeq

It will now be convenient to consider the vacua and mass matrices on a
case-by-case basis.

\subsubsection{\texorpdfstring{$N=M=1$}{N=M=1}}

For a single gauge group with a single flavor of hypers, the FI vector
can without loss of generality be taken to be that of
Eq.~\eqref{eq:r1}.
The solution to the vacuum equations \eqref{eq:vac12}, \eqref{eq:vac3}
is
\beq
\rho=\tilde\rho=\sqrt{\frac{\alpha}{|Q|}}, \qquad
\vartheta+\tilde\vartheta=\left(\frac{1-\sign(Q)}{2}\right)\pi,
\label{eq:vacsol_N=M=1}
\eeq
where we have suppressed the flavor index and $\vartheta-\tilde\vartheta$ is
a vacuum modulus. 
The gauge symmetry is spontaneously broken and the photon mass reads
\beq
\calM^2 = 4e^2|Q|\alpha > 0.
\eeq

\subsubsection{\texorpdfstring{$N=M=2$}{N=M=2}}

For the $N=M=2$ case, we have two FI vectors and they can without loss
of generality be taken to be those of Eqs.~\eqref{eq:r1} and
\eqref{eq:r2} and hence the vacuum equation \eqref{eq:vac3} forces
$\rho_A=\tilde\rho_A$ for both flavors.
Changing variables to
\beq
\bz \equiv
\begin{pmatrix}
  2\rho_1^2e^{\i(\vartheta_1+\tilde\vartheta_1)}\\
  2\rho_2^2e^{\i(\vartheta_2+\tilde\vartheta_2)}
\end{pmatrix}^{\rm T},
\eeq
the vacuum solution reads
\beq
\bz = 
\begin{pmatrix}
  \alpha, & 
  \beta + \i\gamma
\end{pmatrix} Q^{-1}.
\label{eq:vacsol_N=M=2}
\eeq
In case of $\gamma=0$, which corresponds to the situation where the
two FI vectors were parallel (proportional to each other) before the
$\SU(2)_R$ rotation to the standard form \eqref{eq:r1},
\eqref{eq:r2}, the solution $\bz$ is real valued and hence
\beq
\vartheta_A + \tilde\vartheta_A = w_A\pi, \qquad
w_A\in\mathbb{Z}, \qquad
\forall A=1,2,
\eeq
correspond to the signs of the solution \eqref{eq:vacsol_N=M=2}.

The determinant of the mass-squared matrix can thus be written as
\beq
\det(\calM^2) &= 16e_1^2e_2^2 (\det Q)^2 \rho_1^2\rho_2^2, 
\label{eq:detMsq_N=M=2}
\eeq
where we recall that $\rho_A=\tilde\rho_A$.
Using now that $|z_A|=2\rho_A^2$ and
\beq
(\det Q)\bz =
\begin{pmatrix}
  \alpha Q_{22} - Q_{21}(\beta + \i\gamma)\\
  -\alpha Q_{12} + Q_{11}(\beta + \i\gamma)
\end{pmatrix},
\eeq
we have
\beq
\det(\calM^2) = 4e_1^2e_2^2
\sqrt{\left(\gamma^2 Q_{11}^2 + (\alpha Q_{12} - \beta Q_{11})^2\right)
  \left(\gamma^2Q_{21}^2 + (\alpha Q_{22} - \beta Q_{21})^2\right)},
\label{eq:detMsqN=M=2}
\eeq
and for $\gamma=0$, this can clearly vanish even if $\det Q\neq 0$.
Of course it is a special case of aligned FI vectors, which in turn
align with the charge lattice. 

Considering the exceptional case of $\gamma=0$, $\det Q\neq 0$ for
which the gauge symmetry is not completely broken, we see from
Eq.~\eqref{eq:detMsqN=M=2} that for either flavor $A$, we have
\beq
\frac{Q_{A1}}{Q_{A2}} = \frac{\alpha}{\beta},
\label{eq:N=M=2_special_unbroken_cond}
\eeq
which in turn must be rational for the elements of $Q$ to be able to
take such values.
The vacuum solution in this case ($\gamma=0$, $\det Q\neq 0$), reads
\beq
\rho_B = \sqrt{\left|\sum_{A=1}^2\frac{\epsilon_{BA}(\alpha Q_{A2} - \beta Q_{A1})}{2\det Q}\right|},\qquad
e^{\i(\vartheta_A+\tilde\vartheta_A)} = \sign\left(\sum_{A=1}^2\frac{\epsilon_{BA}(\alpha Q_{A2} - \beta Q_{A1})}{2\det Q}\right).
\eeq
Now if the contrived alignment between the FI vector is in turn
aligned with the charge lattice according to
Eq.~\eqref{eq:N=M=2_special_unbroken_cond}, the symmetry is not
completely broken spontaneously 
\begin{align}
  \rho_1 &= 0,\non
  \rho_2 &= \sqrt{\left|\frac{\beta}{Q_{22}}\right|}, \qquad
  e^{\i(\vartheta_A+\tilde\vartheta_A)} = \sign\left(\frac{\beta}{Q_{22}}\right).
\end{align}

In summary we have the condition for broken gauge symmetry: $\det Q\neq 0$
which is also necessary for invertibility of $Q$ for the vacuum
solution.
If $\gamma=0$, we have the additional conditions:
$\frac{Q_{A1}}{Q_{A2}}\neq\frac{\alpha}{\beta}$, $\forall A=1,2$.

\subsubsection{\texorpdfstring{$N=M=3$}{N=M=3}}

Although we can simplify the first 2 FI vectors, the third vector is
generic even after the $\SU(2)_R$ rotations. Hence, we generically
have $\br_1=(\alpha>0,0,0)$, $\br_2=(\beta,\gamma\geq 0,0)$ and
$\br_3=(\delta,\kappa,\eta)$.
Changing variables to
\beq
\bz \equiv
\begin{pmatrix}
  2\rho_1\tilde\rho_1 e^{\i(\vartheta_1+\tilde\vartheta_1)}\\
  2\rho_2\tilde\rho_2 e^{\i(\vartheta_2+\tilde\vartheta_2)}\\
  2\rho_3\tilde\rho_3 e^{\i(\vartheta_3+\tilde\vartheta_3)}
\end{pmatrix}^{\rm T}\in\mathbb{C}^3, \qquad
\by \equiv
\begin{pmatrix}
  \rho_1^2 - \tilde\rho_1^2\\
  \rho_2^2 - \tilde\rho_2^2\\
  \rho_3^2 - \tilde\rho_3^2
\end{pmatrix}^{\rm T}\in\mathbb{R}^3,
\eeq
the vacuum equations \eqref{eq:vac12} and \eqref{eq:vac3} are solved
by 
\beq
\bz = 
\begin{pmatrix}
  \alpha, &
  \beta + \i\gamma, &
  \delta + \i\kappa
\end{pmatrix} Q^{-1}, \qquad
\by = 
\begin{pmatrix}
  0, &
  0, &
  \eta
\end{pmatrix} Q^{-1}.
\eeq
The determinant of the mass-squared matrix of the photons for this
case is again given by Eq.~\eqref{eq:detMsqN=M}, which we now can
write as 
\begin{align}
  \det(\mathcal{M}^2) &= 8e_1^2e_2^2e_3^2(\det Q)^2
  \prod_{A=1}^3(\rho_A^2 + \tilde\rho_A^2) \non
  &= 8e_1^2e_2^2e_3^2(\det Q)^2 \prod_{A=1}^3 \sqrt{|z_A|^2+y_A^2} \non
  &= 8e_1^2e_2^2e_3^2|\det Q|^{-1} \prod_{A=1}^3\sqrt{
    \left(\alpha\Upsilon_{A1} - \beta\Upsilon_{A2} + \delta\Upsilon_{A3}\right)^2
    +\left(\gamma\Upsilon_{A2} - \kappa\Upsilon_{A3}\right)^2
    +\eta^2\Upsilon_{A3}^2},
  \label{eq:detMsqN=M=3}
\end{align}
where $\Upsilon_{Aa}\equiv\det Q^{Aa}$ is the minor determinant of $Q$
with the $A$th row and the $a$th column removed. 
Note that $\eta\Upsilon_{A3}\neq 0,\forall A$ and $\det Q\neq 0$ are
sufficient conditions for ensuring that the gauge symmetry is
completely broken (spontaneously).

If on the other hand, $\eta=0$ or one of the three $\Upsilon_{A3}=0$,
it is possible to find accidental solutions with a vanishing
eigenvalue of the mass-squared matrix even though $\det Q\neq 0$. 
Starting with $\eta=0$, we have the equations giving rise to a
vanishing determinant of the mass-squared matrix
\beq
\alpha\Upsilon_{A1} + \delta\Upsilon_{A3} = \beta\Upsilon_{A2},\qquad
\gamma\Upsilon_{A2} = \kappa\Upsilon_{A3},
\eeq
and choosing the flavor giving rise to a vanishing eigenvalue to be
$A=3$, we can write the solution as
\beq
\bz =
\frac{1}{\det Q}
\begin{pmatrix}
  \alpha\Upsilon_{11} - (\beta+\i\gamma)\Upsilon_{12} + (\delta+\i\kappa)\Upsilon_{13}\\
  -\alpha\Upsilon_{21} + (\beta+\i\gamma)\Upsilon_{22} - (\delta+\i\kappa)\Upsilon_{23}\\
  0
  \end{pmatrix}^{\rm T}, \qquad
\by = \mathbf{0}.
\eeq
The other exceptional case which could result in a partially unbroken
gauge symmetry is $\Upsilon_{A3}=0$, $\eta\neq 0$ and $\gamma=0$. We
will again choose the guilty flavor to be $A=3$. The fact that
$\gamma$ must vanish as well in this exceptional case is because the
requirement $\det Q\neq 0$ does not allow us to have $\Upsilon_{32}=0$
for $\Upsilon_{33}=0$. We thus have 
\beq
\alpha\Upsilon_{31} = \beta\Upsilon_{32},
\eeq
and the vacuum solution is
\beq
\bz =
\frac{1}{\det Q}
\begin{pmatrix}
  \alpha\Upsilon_{11} - \beta\Upsilon_{12} + (\delta+\i\kappa)\Upsilon_{13}\\
  -\alpha\Upsilon_{21} + \beta\Upsilon_{22} - (\delta+\i\kappa)\Upsilon_{23}\\
  0
\end{pmatrix}^{\rm T}, \qquad
\by = \frac{1}{\det Q}
\begin{pmatrix}
  \eta\Upsilon_{13}\\
  -\eta\Upsilon_{23}\\
  0
\end{pmatrix}^{\rm T}.
\eeq

In summary we have the condition for broken gauge symmetry
$\det Q\neq 0$, which is also a necessity for the invertibility of
the charge matrix $Q$.
If, however, $\eta=0$ or $\gamma=0$, a conspiracy amongst the FI
parameters and the charges can occur so that the gauge symmetry is
partially unbroken.

\section{1/2-BPS strings}

In this section, we will consider 1/2-BPS strings (states), which by
the nature of supersymmetry algebra are always parallel strings.
Thus, without loss of generality, we can take the string to point in
the $x^3$ direction and hence the nontrivial behavior is all contained
in the $(x^1,x^2)$-plane.
Next, we will consider the supersymmetry projections that will spell
out the BPS equations.

\subsection{Supersymmetry projections}\label{sec:SUSYproj}

A fermion field is described by a
$2^{\left\lfloor\frac{d}{2}\right\rfloor}$-dimensional spinor, which is a
$2^{\left\lfloor\frac{d}{2}\right\rfloor}$-dimensional representation of an
SU(2) field.
In particular, in $d+1=6$ dimensions, a fermion field has 8
components, whereas in $d+1=4$ dimensions (like nature) it has only 4
components.
This makes it convenient to use $\mathcal{N}=1$ supersymmetry in
$d+1=6$ dimensions, which naturally possesses 8 supercharges (the
eight components of the fermionic spinor), as a shortcut to
$\mathcal{N}=2$ supersymmetry in $d+1=4$ dimension (which has
$(\mathcal{N}=2)\times 4=8$ supercharges as well).
An appropriate dimensional reduction gives the field content and field
equations in $d+1=4$ dimensions, see e.g.~Ref.~\cite{Eto:2005sw}.
In this section, we will use the invariance of $\mathcal{N}=1$
supersymmetry in $d+1=6$ dimensions as a method to obtain the BPS
equations for the $\mathcal{N}=2$ theory \eqref{eq:L} in $d+1=4$
dimensions. 

The supersymmetry transformation is realized by means of an SU(2)
Majorana-Weyl spinor, $\epsilon_i$, with $i=1,2$ that satisfies 
\begin{align}
  \sum_{j}\Gamma_{ij}^{012345}\epsilon_j &= \epsilon_i,\\
  B\bar\epsilon^i &= \sum_{j}\varepsilon^{ij}\epsilon_j,
\end{align}
where we have defined the 6-dimensional ``$\gamma^5$'' as
\beq
\Gamma^{012345} \equiv -\Gamma^0\Gamma^1\Gamma^2\Gamma^3\Gamma^4\Gamma^5,
\eeq
and $B$ can be defined as
\beq
B = \Gamma^{012345}\Gamma^3\Gamma^5.
\eeq
In general
\beq
\Gamma^{K_1\cdots K_n} \equiv \frac{1}{n!}\Gamma^{[K_1}\cdots\Gamma^{K_n]},
\eeq
is the normalized totally antisymmetric product of $n$ $\Gamma$
matrices and $K_1,\ldots,K_n$ are 6-dimen\-sional spacetime indices. 
Because of the Clifford algebra
\beq
\left\{\Gamma^{K_1},\Gamma^{K_2}\right\} = 2\eta^{K_1K_2},
\eeq
the anti-symmetrized product of $\Gamma$ matrices, for $K_1$ and $K_2$
different from each other ($K_1\neq K_2$), is simply the product of the
two Gamma matrices, and so on for higher $n$. 
For more details on the $\Gamma$ matrices and supersymmetry in higher
dimensions, see Appendix B of Polchinski's book
\cite{Polchinsky:1998}.  
The Majorana-Weyl spinor with eight supercharges is given by
\begin{align}
  \epsilon_1 &= (p,0,0,q,0,r,s,0),\non
  \epsilon_2 &= (-\bar{q},0,0,\bar{p},0,-\bar{s},\bar{r},0),
  \label{eq:epsilon}
\end{align}
where $p,q,r,s$ are complex Grassmann variables. 

As mentioned above, we can without loss of generality choose the
string to point in the $x^3$ direction, for which the appropriate
combination of $\Gamma$ matrices is $\Gamma^{12}=\Gamma^1\Gamma^2$.
The most general 1/2-BPS supersymmetry projection can thus be written as
\beq
P_{ij}^{\bm} = \Gamma^{12}\otimes(\i\bm\cdot\bsigma)_{ij},
\label{eq:SUSYproj}
\eeq
with $\bm=(m_1,m_2,m_3)$ a unit 3-vector ($\bm\cdot\bm=1$) and
$\bsigma$ a 3-vector of the three Pauli SU(2) matrices. 

The invariance
\beq
\sum_jP_{ij}^{\bm}\epsilon_j = \epsilon_i,
\eeq
yields the following relations among the supercharges
\beq
p = \i m_- \bar{r} - \i m_3 s, \qquad
q = m_- \bar{s} + \i m_3 r, \qquad
m_\pm \equiv m_1 \pm\i m_2,
\eeq
which confirms that there are 4 conserved supercharges and hence it is
a 1/2-BPS projection, as promised. 
Notice that the unit vector $\bm$ rotates the combinations of $r$ and
$s$ in the 1/2-BPS supersymmetry relations between the supercharges.

\subsection{BPS equations}

The most general SUSY projection \eqref{eq:SUSYproj} is all
that is needed for generating the 1/2-BPS equations that we want to
study.
Indeed, we consider the supersymmetry transformations of the gaugino
and the hyperino -- which are fermion fields belonging to the gauge
and the scalar sectors -- by the Majorana-Weyl 
spinor $\epsilon$ in 6 dimensions, which read \cite{Eto:2005sw} 
\begin{align}
  \delta_\epsilon\lambda^a &= \frac12\Gamma^{K_1K_2}F_{K_1K_2}^a\epsilon
  +\i\bY_a\cdot\bsigma\epsilon,\\
\delta_\epsilon\psi_A &= \Gamma^{K} D_{K}\Phi_A^{\rm T} \sigma^2\epsilon,
\end{align}
where $a=1,2,\ldots,N$, is the gauge group index for $\U(1)_a$,
$A=1,2,\ldots,M$, is the flavor index and $\bY_a$ is an $\SU(2)_R$
triplet of D and F terms.  
We will now trivialize the fields in the extra-dimensional spatial
directions $K=4,5$ as well as in the spatial direction $K=3$ ($\mu=3$)
(due to translational invariance in this direction).
Forcing the above two fermion transformations to vanish, we have
\begin{align}
  \Gamma^{12}F_{12}^a\epsilon + \i\bY_a\cdot\bsigma\epsilon &= 0,\\
  \Gamma^1\left(D_1\Phi_A^{\rm T} + \Gamma^{12}D_2\Phi_A^{\rm T}\right)\sigma^2\epsilon &= 0.
\end{align}
Utilizing the SUSY projection \eqref{eq:SUSYproj}, we get
\begin{align}
  -\i F_{12}^a(\bm\cdot\bsigma)\epsilon + \i\bY_a\cdot\bsigma\epsilon = 0,\\
  \Gamma^1\left(D_1\Phi_A^{\rm T} -\i D_2\Phi_A^{\rm T}\sigma^2(\bm\cdot\bsigma)\sigma^2\right)\sigma^2\epsilon = 0.
\end{align}
Using that $\sigma^2(\bm\cdot\bsigma)\sigma^2=-(\bm\cdot\bsigma)^{\rm T}$,
we can readily transpose the second equation, obtaining the set of
1/2-BPS equations
\begin{align}
  (\bm\cdot\bsigma) F_{12}^a &= e_a^2\bsigma\cdot
  \bigg(\sum_A Q_{Aa}\Phi_A^\dag\bsigma\Phi_A - \br_a\bigg),\label{eq:BPS1}\\
  D_1\Phi_A + \i(\bm\cdot\bsigma) D_2\Phi_A &= 0,\label{eq:BPS2}
\end{align}
where we have inserted the D- and F-terms $\bY_a$ of Eq.~\eqref{eq:Y}.

\subsection{Bogomol'nyi bound}

By knowing the BPS equations \eqref{eq:BPS1}, \eqref{eq:BPS2} as well
as the Lagrangian \eqref{eq:Lreduced}, it is fairly straightforward to
write down the Bogomol'nyi bound for the theory
\begin{align}
\calE &=
\sum_A\left|D_1\Phi_A + \i\bm\cdot\bsigma D_2\Phi_A\right|^2
+\sum_{a,\alpha}\frac{1}{2e_a^2}\bigg[
m^\alpha F_{12}^a - e_a^2\Big(\sum_A\Phi_A^\dag\sigma^\alpha\Phi_A Q_{Aa} - r_a^\alpha\Big)
\bigg]^2 \non
&\phantom{=\ }
- \sum_a \bm\cdot \br_a F_{12}^a,
\label{eq:BPSenergy}
\end{align}
where we have assumed that no field is dependent on $x^3$, that
all $A_3^a=0$ vanish and we have used the fact that
\beq
[D_1,D_2]\Phi_A = \i\sum_a Q_{Aa} F_{12}^a \Phi_A.
\eeq
Since the first two terms of Eq.~\eqref{eq:BPSenergy} are positive
semi-definite, the total energy in the $(x^1,x^2)$-plane, which is the
string tension, is bounded from below.
Integrating over the $(x^1,x^2)$-plane thus gives the Bogomol'nyi bound
\beq
T = \int_{\mathbb{R}^2} \calE\;\d{x}^1\d{x}^2
\geq 2\pi\sum_a \bm\cdot\br_a k^a,
\eeq
where we have defined the magnetic fluxes
\beq
k^a \equiv -\frac{1}{2\pi}\int_{\mathbb{R}^2} F_{12}^a\;\d{x}^1\d{x}^2.
\eeq
We have seen explicitly in Sec.~\ref{sec:SUSYproj} that the choice of
$\bm$ corresponds to the choice of SUSY projection.
The most stringent bound appears when $\bm$ is parallel with
$\br_a k^a$ in $\SU(2)_R$ space, which makes it an obvious choice for
the SUSY projection, which thus reads
\beq
\bm = \sum_a\frac{\br_a k^a}{\sqrt{\sum_{b,c}\br_b\cdot\br_c k^b k^c}},
\label{eq:1/2BPS_msol}
\eeq
and is normalized to unit length.

The Bogomol'nyi bound can thus be written as
\beq
T \geq T_{\rm BPS} \equiv 2\pi\sqrt{\sum_{a,b}\br_a\cdot\br_b k^a k^b}
= 2\pi\left|\sum_a\br_a k^a\right|.
\label{eq:1/2BPS_Bogomol'nyi_bound}
\eeq
We thus see that the minimal string tension depends not only on the FI
vectors, but also on the magnetic fluxes and which gauge groups they
are turned on in.
Since the BPS tension (i.e.~the tension given by the saturated
Bogomol'nyi bound) is given by the length of the vector
\beq
\sum_a\br_ak^a,
\label{eq:rkvec}
\eeq
the bound can only vanish if this vector is a null vector, viz.~a
vector of length $0$.
$\bm$ on the other hand is, by definition, a unit 3-vector and hence
cannot be a null vector.
In the case that the vector \eqref{eq:rkvec} vanishes, $\bm$ cannot be
determined by Eq.~\eqref{eq:1/2BPS_msol} but can be chosen freely.

\subsection{Vortex and constraint equations}

In this section we will study the conditions for the existence of
solutions to the 1/2-BPS equations, which are nontrivial due to the
fact that Eq.~\eqref{eq:BPS1} has three components for each gauge
group.
This generally gives us one vortex equation and two
constraint equations.

First, analogously to rotating the FI vectors, we will use a more
convenient basis for the vortex fields, which is not the $\Phi$ basis,
but is defined as
\beq
\Phi_A = U\Psi_A = U
\begin{pmatrix}
  \psi_A\\
  \tilde\psi_A^*
\end{pmatrix},
\eeq
with the $\SU(2)_R$ transformation
\beq
U = \sqrt{\frac{1+m_3}{2}}
\begin{pmatrix}
  1 & \frac{m_3 - 1}{m_+}\\
  \frac{1 - m_3}{m_-} & 1
\end{pmatrix},
\eeq
which is readily checked to have determinant one.
We call this the $\Psi$ basis. 
This matrix diagonalizes all $\bm$ vectors into
$\sigma^3$.\footnote{Strictly speaking, we should take the limiting
  value of $U$ for $\bm=(0,0,\pm 1)$, i.e.~$\lim_{m_3\to\pm 1}U$.}
Notice that if $\bm=(0,0,1)$ the diagonalization matrix $U$ above, is
just the unit two-matrix.

The BPS equations \eqref{eq:BPS1} and \eqref{eq:BPS2} can thus be put
in the form
\begin{align}
  F_{12}^a &= e_a^2\Big(\sum_AQ_{Aa}\Psi_A^\dag\sigma^3\Psi_A - \bm\cdot\br_a\Big),\label{eq:BPS1_Psi}\\
  D_1\Psi_A + \i\sigma^3D_2\Psi_A &= 0,\label{eq:BPS2_Psi}
\end{align}
which we shall coin the vortex equations and
\beq
\sum_A \Phi_A^\dag\bar{\bell}\cdot\bsigma\Phi_A Q_{Aa} = \bar{\bell}\cdot\br_a,
\label{eq:constraint_Phi_basis}
\eeq
are the constraint equations, where we have defined the $\SU(2)_R$
complex vector that is orthonormal to $\bm$, (i.e.~$\bm\cdot\bar{\bell}=0$):
\beq
\bar{\bell} \equiv \frac{\bell_1 - \i\bell_2}{\sqrt{2}}
= \frac{1}{\sqrt{2}m_+}
\begin{pmatrix}
  m_1m_3 + \i m_2\\
  m_2m_3 - \i m_1\\
  -m_+m_-
\end{pmatrix}.
\label{eq:bell}
\eeq
Changing to the $\Psi$ basis, the constraint equation reads
\beq
\frac{1}{\sqrt{2}}\sum_A \Psi_A^\dag\sigma^-\Psi_A Q_{Aa} =
\sqrt{2}\sum_A \psi_A\tilde\psi_A Q_{Aa} =
\bar{\bell}\cdot\br_a,
\label{eq:constraint_psi}
\eeq
which is valid for all cases.\footnote{Like for $U$, $\bar\bell$ should
  be taken as the limiting value for $m_3=\pm1$,
  i.e.~$\lim_{m_3\to\pm 1}\bar\bell$, which for $m_3\to 1$ yields
  $\bar\bell=(1,-\i,0)/\sqrt{2}$.
  However, the limit $m_3\to-1$ is ambiguous and this introduces a
  spurious phase $\alpha$,
  i.e.~$\bm=(\sqrt{1-m_3^2}\cos\alpha,\sqrt{1-m_3^2}\sin\alpha,m_3)$,
  which survives the limit of $m_3\to -1$.
  Noticing that $m_3\to -1$ corresponds simply to swapping $\psi_A$
  with $\tilde\psi_A^*$, we can fix the ambiguous phase $\alpha$ so
  the limit reads
  $\lim_{m_3\to-1}\bar\bell=(1,\i,0)/\sqrt{2}$. 
}
It will prove convenient to rewrite the above equation into an
equation with a free flavor index $A$, instead of a free gauge index
$a$:
\beq
\psi_A\tilde\psi_A = \frac{1}{\sqrt{2}}\sum_a
\bar\bell\cdot\br_a Q_{aA}^{-1}
=\frac{1}{\sqrt{2}}\bar\bell\cdot\bOmega_A,
\label{eq:constraint_psi2}
\eeq
where in the last equality we have defined
\beq
\bOmega_A \equiv \sum_a\br_a Q^{-1}_{aA}.
\label{eq:Omega_def}
\eeq

In components, the vortex equations \eqref{eq:BPS1_Psi} and
\eqref{eq:BPS2_Psi} read
\begin{align}
  D_{\bar{z}}\psi_A &= 0,\label{eq:BPS1_psi}\\
  D_{\bar{z}}\tilde\psi_A &= 0,\label{eq:BPS2_psi}\\
  2\i F_{\bar{z}z}^a &= e_a^2\bigg(
  \sum_A |\psi_A|^2Q_{Aa}
  -\sum_A |\tilde\psi_A|^2Q_{Aa} - \bm\cdot\br_a\bigg),\label{eq:BPS3_psi}
\end{align}
where we have switched to complex coordinates in the
$(x^1,x^2)$-plane, i.e.~$z=x^1+\i x^2$.
Although the two first BPS equations look identical, the field
$\psi_A$ has charges $Q_{Aa}$, whereas $\tilde\psi_A$ is
an antifundamental field and thus has the charges $-Q_{Aa}$.
This sign has severe consequences for the existence of solutions.

It will prove instructive to flesh out the self-dual equation
\eqref{eq:BPS1_psi} as
\beq
D_{\bar{z}}\psi_A =
\p_{\bar{z}}\psi_A + \i\sum_a Q_{Aa} A_{\bar{z}}^a\psi_A = 0, \qquad
\forall A.
\eeq
We can solve for the gauge fields, obtaining
\beq
A_{\bar{z}}^a = \i\sum_A Q^{-1}_{aA}\p_{\bar{z}}\log\psi_A.
\label{eq:Azbar_psi}
\eeq
We may split the field $\psi_A$ into a holomorphic and a nonanalytic
field
\beq
\psi_A(z,\bar{z}) = \varsigma^{-1}_A(z,\bar{z}) h_A(z).
\eeq
The gauge field does not depend on $h_A(z)$ and thus reads
\beq
A_{\bar{z}}^a = -\i\sum_AQ^{-1}_{aA}\p_{\bar{z}}\log\varsigma_A.
\label{eq:Azbar_psi_varsigma}
\eeq
The field strength tensor can now be written as
\beq
F_{12}^a = 2\i F_{\bar{z}z}^a =
-2\sum_{A}Q^{-1}_{aA}\p_{\bar{z}}\p_z\log|\varsigma_A|^2.
\eeq
Integrating over the $(x^1,x^2)$-plane, we get
\begin{align}
k^a = -\frac{1}{2\pi}\int_{\mathbb{R}^2} F_{12}^a \;\d{x}^1\d{x}^2
&= \frac{1}{\pi}\sum_AQ^{-1}_{aA}
\int_{\mathbb{R}^2} \p_{\bar{z}}\p_z\log|\varsigma_A|^2\;\d{x}^1\d{x}^2\non
&= \frac{1}{2\pi\i}\sum_AQ^{-1}_{aA}
\oint_{\p\mathbb{R}^2} \p_z\log|\varsigma_A|^2\;\d{z},
\label{eq:ka_integral}
\end{align}
where we have used Green's theorem.

In order to know the asymptotic behavior of $\varsigma_A$, we need to
transform the vacuum solution from the $\Phi$ basis to the $\Psi$
basis
\begin{align}
|\langle\psi_A\rangle|^2 = \frac12\left(
(1+m_3)|\langle\phi_A\rangle|^2
+(1-m_3)|\langle\tilde\phi_A\rangle|^2
+m_+\langle\phi_A\tilde\phi_A\rangle
+m_-\langle\phi_A^*\tilde\phi_A^*\rangle
\right),\\
|\langle\tilde\psi_A\rangle|^2 = \frac12\left(
(1-m_3)|\langle\phi_A\rangle|^2
+(1+m_3)|\langle\tilde\phi_A\rangle|^2
-m_+\langle\phi_A\tilde\phi_A\rangle
-m_-\langle\phi_A^*\tilde\phi_A^*\rangle
\right).
\end{align}
Using now the vacuum solution \eqref{eq:general_vac_sol}, we can 
write the vacuum in the $\Psi$ basis as
\begin{align}
  |\langle\psi_A\rangle|^2 = \frac12\left(
  \sqrt{y_A^2 + |z_A|^2} + m_3 y_A + \frac{m_+}{2}z_A^* + \frac{m_-}{2}z_A
  \right),\\
  |\langle\tilde\psi_A\rangle|^2 = \frac12\left(
  \sqrt{y_A^2 + |z_A|^2} - m_3 y_A - \frac{m_+}{2}z_A^* - \frac{m_-}{2}z_A
  \right).
\end{align}
Utilizing the useful relations \eqref{eq:useful_vac_rels} and the
definition \eqref{eq:Omega_def}, we arrive at
\beq
|\langle\psi_A\rangle|^2 = \frac12\left(
|\bOmega_A| + \bm\cdot\bOmega_A
\right),\qquad
|\langle\tilde\psi_A\rangle|^2 = \frac12\left(
|\bOmega_A| - \bm\cdot\bOmega_A
\right),
\label{eq:psiA_vac_mvec}
\eeq
with $|\bOmega_A|\equiv\sqrt{\bOmega_A\cdot\bOmega_A}$.
Clearly, the sum
\beq
|\langle\psi_A\rangle|^2
+|\langle\tilde\psi_A\rangle|^2
= |\bOmega_A| \neq 0,
\label{eq:bOmega_cannot_vanish}
\eeq
is always nonvanishing whereas if $\bOmega_A$ is parallel
(anti-parallel) with $\bm$ then $|\langle\psi_A\rangle|^2>0$
($|\langle\tilde\psi_A\rangle|^2>0$) is nonvanishing.
Therefore, if one of the two VEVs $|\langle\psi_A\rangle|^2$,
$|\langle\tilde\psi_A\rangle|^2$ vanishes, the other must be
nonvanishing.

We will now assume that $|\langle\psi_A\rangle|^2>0$; therefore, the
boundary condition on 
$\lim_{|z|\to\infty}|\psi_A|\to|\langle\psi_A\rangle|$ leads to
\beq
\lim_{|z|\to\infty}\left|\frac{\varsigma_A(z,\bar{z})}{h_A(z)}\right|^2
  = \frac{1}{|\langle\psi_A\rangle|^2}.
\eeq
Because there exists no holomorphic U(1) bundle of negative degree, we
must have
\beq
h_A(z) = \sum_{p=0}^{n_A} h_{A,p} z^p,
\eeq
with $h_{A,p}\in\mathbb{C}$ being constants, therefore
\beq
\lim_{|z|\to\infty}|\varsigma_A(z,\bar{z})|^2 =
\frac{|z|^{2n_A}}{|\langle\psi_A\rangle|^2}
=\frac{z^{n_A}\bar{z}^{n_A}}{|\langle\psi_A\rangle|^2},
\eeq
and in turn by Eq.~\eqref{eq:ka_integral}, we obtain
\beq
k^a =
\frac{1}{2\pi\i}\sum_{A}Q^{-1}_{aA}\oint\frac{n_A}{z}\;\d{z}
= \sum_{A}Q^{-1}_{aA}n_A,
\eeq
where $n_A$ is the winding number of the flavor $A$.
Notice that $k^a$ are not necessarily integers, but must be rational
numbers. 
Solving for the winding numbers, we get
\beq
n_A = \sum_a Q_{Aa}k^a.
\eeq

Now we can repeat the calculation for $\tilde\psi_A$ which is
completely analogous, except that its charge is $-Q_{aA}$ and we now 
assume that $|\langle\tilde\psi_A\rangle|^2>0$, hence 
\beq
\tilde{n}_A = -\sum_a Q_{Aa}k^a.
\eeq
Since it is impossible to have holomorphic U(1) bundles of negative
degrees, neither $n_A$ nor $\tilde{n}_A$ can be negative.
Unfortunately, as they are determined by the same expression, a
non-negative $\sum_aQ_{Aa}k^a$ will inevitably turn
on a positive $n_A$ and a negative $\tilde{n}_A$ or \emph{vice versa}. 
The only way out is the following lemma.

\begin{lemma}\label{lemma:1}
  For winding flavors of 1/2-BPS vortices in the theory
  \eqref{eq:Lreduced}, i.e.~$n_A\neq 0$, either
  $\langle\tilde\psi_A\rangle=0$ or $\langle\psi_A\rangle=0$, since if
  nonvanishing they will possess winding and they cannot both be
  winding since that would make one of them winding with a negative
  holomorphic function and in turn be singular. 
\end{lemma}
\emph{Proof}: Asymptotically the behavior of the scalar fields is
\begin{align}
  \lim_{|z|\to\infty}\psi_A &= |\langle\psi_A\rangle| e^{\i n_A\theta},\\
  \lim_{|z|\to\infty}\tilde\psi_A &= |\langle\tilde\psi_A\rangle| e^{-\i n_A\theta},
\end{align}
where we have used that $\tilde{n}_A=-n_A$.
Since neither field can have negative winding,
$\langle\tilde\psi_A\rangle$ must vanish if $n_A>0$ and contrarily
$\langle\psi_A\rangle$ must vanish if $n_A<0$.\hfill$\square$

\begin{corollary}\label{coro:Omega}
  Since either $\langle\tilde\psi_A\rangle=0$ or
  $\langle\psi_A\rangle=0$ for all winding flavors ($n_A\neq 0$) by
  lemma \ref{lemma:1}, the vector $\bm$ must correspondingly be
  parallel or anti-parallel with $\bOmega_A$. 
\end{corollary}
\emph{Proof}:
By inspection of Eq.~\eqref{eq:psiA_vac_mvec} the corollary
immediately follows.
\hfill$\square$

We will now contemplate the solutions to the constraint
Eq.~\eqref{eq:constraint_psi2}. 
Considering first asymptotic distances, we only need to know the
scalar fields VEVs $\langle\psi_A\rangle$,
$\langle\tilde\psi_A\rangle$, the charge matrix $Q$, the FI vectors
$\br_a$ and the vector $\bm$ which is given by
Eq.~\eqref{eq:1/2BPS_msol} in terms of $k^a$ and $\br_a$.
Each winding flavor must have $\langle\psi_A\rangle=0$ or
$\langle\tilde\psi_A\rangle=0$ and for that flavor, the right-hand
side of Eq.~\eqref{eq:constraint_psi2} must accordingly vanish; this
puts constraints of the allowed fluxes $k^a$ for given FI vectors and
charge matrix.
Considering now coming in from infinity to non-asymptotic distances.
The field that has a vanishing VEV, say $\tilde\psi_A$, must remain
strictly vanishing throughout the entire $(x^1,x^2)$-plane or else there is
no smooth way to satisfy the constraint
Eq.~\eqref{eq:constraint_psi2}. 
The other field -- the winding field -- vanishes only at the $n_A$
vortex centers, which in turn have Lebesgue measure zero.

These considerations can be summarized in the following proposition:
\begin{prop}\label{prop:1}
The 1/2-BPS vortex solutions in the theory \eqref{eq:Lreduced} which
solve the constraint Eq.~\eqref{eq:constraint_psi2} and is compatible
with the result of Lemma \ref{lemma:1} can be classified into the
following categories, based on whether $\bar\bell\cdot\br_a$ vanishes
for all $a$ or not:  
\begin{itemize}
\item\emph{Type A solutions}: Either $\psi_A\equiv 0$ or
  $\tilde\psi_A\equiv 0$ throughout $\mathbb{R}^2$ and hence
  $\sum_a\bar\bell\cdot\br_a Q^{-1}_{aA}=0$, for
  every flavor $A$. 
  This allows all flavors to have nonvanishing winding numbers:
  $n_A\neq 0, \forall A$.
\item\emph{Type B solutions}: All winding flavors have either
  $\psi_A\equiv 0$ or $\tilde\psi_A\equiv 0$ throughout $\mathbb{R}^2$
  for $A\in S_{\rm WF}$ (i.e.~the set of winding flavors (WF)) for
  which $\sum_a\bar\bell\cdot\br_a Q^{-1}_{aA}=0$, 
  and the remaining ``inert'' flavors are constant solutions obeying
  \beq
  \psi_A\tilde\psi_A = \frac{1}{\sqrt{2}}
  \sum_a\bar\bell\cdot\br_a Q^{-1}_{aA},\qquad
  A\notin S_{\rm WF},
  \eeq
  which must be nonvanishing for at least one flavor
  $A\notin S_{\rm WF}$ because we demand that
  \beq
  \bar\bell\cdot\br_a \neq 0,
  \eeq
  for at least one gauge group $a$. 
  Although the product $\psi_A\tilde\psi_A$ for a given inert flavor
  $A\notin S_{\rm WF}$ is constant, the individual fields $\psi_A$ and
  $\tilde\psi_A$ do not have to be constant, as they may be coupled to
  nontrivial magnetic fluxes. 
\end{itemize}
\end{prop}

\subsubsection{Type A solutions}

For the type A solutions, we must have
\beq
\psi_A =
\frac{(1+m_3)\phi_A + m_-\tilde\phi_A^*}{\sqrt{2(1+m_3)}}
= 0, \qquad\textrm{or}\qquad
\tilde\psi_A =
\frac{(1+m_3)\tilde\phi_A - m_-\phi_A^*}{\sqrt{2(1+m_3)}}
= 0, \qquad\forall A,
\label{eq:typeA_constraint1}
\eeq
which by corollary \ref{coro:Omega} implies that
$\bOmega_A\propto\pm\bm$. 

The constraint
$\sum_a\bar\bell\cdot\br_a Q^{-1}_{aA}=0$
can be reformulated geometrically as
\beq
\bOmega_A = \sum_a\br_a Q^{-1}_{aA} \propto \bm \propto
\sum_a\br_a k_a.
\label{eq:typeA_constraint2}
\eeq
One could be mislead to think that this relates the magnetic fluxes
$k_a$ and the charges in the charge matrix $Q_{Aa}$.
However, the geometric constraint is really a constraint on $\br_a$ as
stated in the following theorem.

\begin{theorem}\label{thm:1}
The constraint \eqref{eq:constraint_psi2} arising in the theory
\eqref{eq:Lreduced}, in the case of 1/2-BPS type A solutions
(i.e.~$\psi_A=0$ or $\tilde\psi_A=0$ for each $A$), implies that $\bm$
must be proportional to $\br_1=(\alpha,0,0)$, where the latter
direction is a choice of fixing  $\SU(2)_R$ rotational freedom. Hence,
in this case $\bm$ must be $(\pm1,0,0)$. 
\end{theorem}
\emph{Proof}:
Using the constraint equation in the formulation of
Eq.~\eqref{eq:typeA_constraint2}, we must have
\beq
\bOmega_A = \sum_a\br_a Q^{-1}_{aA} \propto \bm,
\eeq
which has two solutions: Either all the vectors $V_a$
\beq
Q^{-1}_{aA} =
\begin{pmatrix}
  V_{1}\\
  V_{2}\\
  \vdots\\
  V_{N}
\end{pmatrix},
\eeq
must be parallel or all the vectors $\br_a$ must be parallel.
If the vectors $V_a$ are parallel it implies that $\det Q^{-1}=0$, which
in turn implies that $Q$ does not exist, and so we must discard this 
solution.
The only remaining solution is that all $\br_a$ are parallel for all
$a$.
Since we have -- without loss of generality -- rotated $\br_1$ to
$\br_1=(\alpha>0,0,0,)$, we must have
\beq
\br_a\propto\br_1, \qquad \forall a.
\eeq
Finally, since $\bm$ is a unit vector proportional to a linear
combination vectors $\br_a$ which are all proportional to
$\br_1$ we have $\bm=(\pm1,0,0)$ and in particular by
Eq.~\eqref{eq:1/2BPS_msol}, we have
\beq
\bm = \bigg(\sign\Big(\sum_a r_a^1 k^a\Big),0,0\bigg),
\label{eq:typeA_m1sign}
\eeq
except for the exceptional case where $\sum_a \br_a k^a=0$ vanishes.
In that case, the sign of the first component of $\bm$ may be chosen
freely. 
\hfill$\square$

This theorem thus immediately leads to the following corollary:
\begin{corollary}\label{coro:2}
  The vacuum of 1/2-BPS type A vortices of the theory
  \eqref{eq:Lreduced} is given by the real solution 
  \beq
  z_A = \sum_a r_a^1 Q^{-1}_{aA},
  \eeq
  and the sign of $m_1z_A$ determines whether the winding field is
  $\psi_A$ (for $m_1z_A>0$) and hence $n_A>0$ or $\tilde\psi_A$ (for
  $m_1z_A<0$) and hence $\tilde{n}_A=-n_A>0$.
  In particular, if $n_A=\sum_a Q_{Aa}k^a$ is nonvanishing, it must
  have the same sign as that of $m_1z_A$. 
\end{corollary}
\emph{Proof}:
For type A solutions, due to theorem \ref{thm:1}, we have
$r_a^2=r_a^3=0$, for all $a$, which by Eq.~\eqref{eq:vac3} implies
that $\rho_A=\tilde\rho_A$ and by Eq.~\eqref{eq:vac12} yields the real
solution for $z_A$
\beq
z_A \equiv 2\rho_A^2 e^{\i(\vartheta_A + \tilde\vartheta_A)} =
\sum_a r_a^1 Q^{-1}_{aA}.
\eeq
Since $\rho_A>0$, the sign of $z_A$ is the sign of the real quantity
$e^{\i(\vartheta_A + \tilde\vartheta_A)}=\pm 1$,
i.e.~$\sign(z_A)=e^{\i(\vartheta_A + \tilde\vartheta_A)}$.
By definition \eqref{eq:Phi_vac}, the VEV of $\Phi$ is determined as
\beq
\langle\Phi\rangle =
e^{\i\tilde\vartheta_A}
\begin{pmatrix}
  \rho_A e^{-\i(\vartheta_A+\tilde\vartheta_A)}\\
  \rho_A
\end{pmatrix}.
\eeq
For $\bm=(\pm1,0,0)$ in accord with theorem \ref{thm:1}, we have by
Eq.~\eqref{eq:typeA_constraint1}, 
\begin{align}
\langle\psi_A\rangle &=
\frac{e^{-\i\vartheta_A}}{\sqrt{2}}
\left(1 + m_1 e^{\i(\vartheta_A+\tilde\vartheta_A)}\right)\rho_A, \\
\langle\tilde\psi_A\rangle &=
\frac{e^{-\i\tilde\vartheta_A}}{\sqrt{2}}
\left(-1 + m_1 e^{\i(\vartheta_A+\tilde\vartheta_A)}\right)\rho_A.
\end{align}
Hence, for positive (negative) $\sign(m_1z_A)$ we
have $\langle\tilde\psi_A\rangle=0$ ($\langle\psi_A\rangle=0$) and the
winding field is $\psi_A$ ($\tilde\psi_A$).
Using now lemma \ref{lemma:1} it follows that if $n_A$ is
nonvanishing, it must have the same sign as $m_1z_A$. 
\hfill$\square$

By using corollary \ref{coro:2} we can relate the magnetic fluxes to
the FI vectors by means of the charge matrix and a set of positive
real numbers.
\begin{theorem}\label{thm:2}
1/2-BPS solutions of type A in the theory \eqref{eq:Lreduced} must
have magnetic fluxes 
\beq
k^a = m_1 \sum_{A,b}Q^{-1}_{aA} c_A^2 r_b^1 Q^{-1}_{bA}, \qquad
c_A^2 z_A \in \mathbb{Z},
\label{eq:thm2}
\eeq
where the $c_A$ are adjusted so all winding numbers
$|n_A|=c_A^2|z_A|\in\mathbb{Z}$ are integers. 
\end{theorem}
\emph{Proof}:
Solutions of type A must have $n_A$ to be either vanishing or of the
same sign as $m_1z_A$ and thus we can write $n_A=c_A^2m_1z_A$, where
$c_A$ are chosen such that $n_A\in\mathbb{Z}$. This yields the
relation
\beq
n_A = c_A^2m_1z_A
= c_A^2m_1\sum_a  r_a^1 Q^{-1}_{aA} \in\mathbb{Z}.
\eeq
$c_A$ may vanish for some $A$, but not for all flavors $A$, since that
corresponds to the trivial solution (no winding numbers in any
fields). 
On the other hand, we have
\beq
n_A = \sum_a Q_{Aa} k^a.
\eeq
Eliminating $n_A$ from the above two equations and multiplying by the
inverse of $Q$, we arrive at Eq.~\eqref{eq:thm2}.
\hfill$\square$

With theorem \ref{thm:2} in hand, we can finally prove that the
somewhat bizarre situation where the vector $\sum_a\br_ak^a$ vanishes,
which in turn leads to a vanishing energy bound, cannot
happen for type A solutions.
\begin{theorem}\label{thm:3}
  1/2-BPS solutions of type A in the theory \eqref{eq:Lreduced} cannot
  have $\sum_a r_a^1k^a=0$, which would imply a vanishing energy bound
  $T\geq T_{\rm BPS}=2\pi\left|\sum_ar_a^1k^a\right|=0$.
\end{theorem}
\emph{Proof}:
The energy bound \eqref{eq:1/2BPS_Bogomol'nyi_bound} simplifies due to
theorem \ref{thm:1} to
\beq
T \geq T_{\rm BPS} = 2\pi\left|\sum_a r_a^1 k^a\right|.
\eeq
Now multiply Eq.~\eqref{eq:thm2} by $r_a^1$ and sum over $a$ to get
\begin{align}
\sum_a r_a^1 k^a
&= m_1\sum_{a,A,b} r_a^1 Q^{-1}_{aA}c_A^2r_b^1Q^{-1}_{bA}\non
&= m_1\sum_A\left(c_A\sum_ar_a^1Q_{aA}^{-1}\right)^2,
\end{align}
where the parenthesis is positive (a square of a real number)
semi-definite and $m_1$ is a sign ($\pm1)$ by theorem \ref{thm:1}.
Thus no cancellation among the terms can happen and the expression can
only vanish if all $c_A=0$, which we cannot allow as that is the
trivial solution (the non-winding vacuum).
\hfill$\square$

\subsubsection{Type B solutions}

For the type B solutions, we split the flavors $A$ into the
set of winding flavors $\tilde{A}\in S_{\rm WF}$ and the rest
$\hat{A}\notin S_{\rm WF}$.
One might think the winding flavors are subject to the same
constraints as in the type A case; this, however, would only be true
if the charge matrix is block diagonal in the winding and non-winding
flavors, respectively, so it does not induce mixing.
Therefore, this type of solutions is nontrivial and the constraints
should be reconsidered carefully. 

For the winding flavors, we shall impose
\beq
n_{\tilde{A}} = \sum_a Q_{\tilde{A}a} k^a \neq 0, \qquad \tilde{A}\in S_{\rm WF},
\label{eq:n_A_WF}
\eeq
and in turn we have by lemma \ref{lemma:1}
\beq
\psi_{\tilde{A}}\tilde\psi_{\tilde{A}} = 0, \qquad \tilde{A}\in S_{\rm WF}.
\eeq
For the non-winding flavors, $\hat{A}\notin S_{\rm WF}$, on the other hand,
we have 
\begin{align}
  2\psi_{\hat{A}}\tilde\psi_{\hat{A}} &=
  \sqrt{2}\sum_a\bar\bell\cdot\br_a Q^{-1}_{a\hat{A}} \non
  &= 
  \left(
  (1+m_3)\phi_{\hat{A}}\tilde\phi_{\hat{A}}
  -\frac{m_-^2}{1+m_3}\phi_{\hat{A}}^*\tilde\phi_{\hat{A}}^*
  -m_-\big(|\phi_{\hat{A}}|^2 - |\tilde\phi_{\hat{A}}|^2\big)
  \right) \non
  &= \sum_a \left(
  \frac{m_1m_3+\i m_2}{m_+}r_a^1
  +\frac{m_2m_3-\i m_1}{m_+}r_a^2
  -m_- r_a^3\right) Q^{-1}_{a\hat{A}}, \qquad \hat{A} \notin S_{\rm WF},
  \label{eq:constraint_nonwinding_flavors}
\end{align}
which may or may not vanish and the magnetic fluxes are determined by 
\beq
k^a = \sum_{\tilde{A}\in S_{\rm WF}} Q^{-1}_{a\tilde{A}} n_{\tilde{A}},
\label{eq:typeB_k}
\eeq
where all the $n_{\tilde{A}}$ in the sum are nonvanishing by
Eq.~\eqref{eq:n_A_WF}.
By definition of the type B vortex solutions, we must have
\beq
\bar\bell\cdot\br_a \neq 0,
\eeq
for some gauge groups $a$, but not necessarily for all $a$. 
This will in turn induce a number of nonvanishing
$\psi_{\hat{A}}\tilde\psi_{\hat{A}}$.
This leads us to the following lemma:
\begin{lemma}\label{lemma:2}
1/2-BPS solutions of type B in the theory \eqref{eq:Lreduced} must
have FI vectors $\br_a$ and charge matrix $Q$ which satisfies
\beq
\sum_a\br_aQ^{-1}_{a\tilde{A}} \propto \bm \propto
\sum_a\sum_{\tilde{B}\in S_{\rm WF}}\br_aQ^{-1}_{a\tilde{B}} n_{\tilde{B}}, \qquad
\forall \tilde{A}\in S_{\rm WF},
\label{eq:lemma2}
\eeq
for nonvanishing windings $n_{\tilde{B}}\neq 0$. 
\end{lemma}
\emph{Proof}:
For type B solutions, we must have
nonvanishing $\sum_a\bar\bell\cdot\br_a Q^{-1}_{aA}\neq 0$ for some
flavors $A$.
These flavors in turn cannot posses nonzero winding numbers, so those
flavors must be $\hat{A}\notin S_{\rm WF}$.
On the contrary, for all winding flavors, we must have
$\psi_{\tilde{A}}\tilde\psi_{\tilde{A}}=0$ by lemma \ref{lemma:1} and
in turn 
\beq
\sum_a\bar\bell\cdot\br_a Q^{-1}_{a\tilde{A}} = 0.
\eeq
Since $\bar\bell$ is a complex vector composed by the only two
orthogonal directions to $\bm$, this means that
\beq
\sum_a \br_a Q^{-1}_{a\tilde{A}} \propto \bm.
\eeq
As the vector $\bm$ is given by Eq.~\eqref{eq:1/2BPS_msol} and is
proportional to $\sum_a\br_ak^a$ and $k^a$ is given by
Eq.~\eqref{eq:typeB_k}, the lemma follows.
\hfill$\square$

\begin{theorem}\label{thm:4}
For 1/2-BPS vortices of type B in the theory \eqref{eq:Lreduced}, the
obvious solution to Eq.~\eqref{eq:lemma2} of lemma \ref{lemma:2} is
that all FI vectors are proportional to each other,
$\br_a\propto\br_1$, but we have to discard that solution because it
is the solution of type A.  
There are two remaining ways to solve Eq.~\eqref{eq:lemma2}:
\begin{itemize}
\item\emph{Type B1 solutions}: There is only 1 winding flavor and the
  remaining flavors are inert.
\item\emph{Type B2 solutions}: The charge matrix is block diagonal,
  such that the winding 
  flavors are all charged under one block of the charge matrix, which
  in turn only turn on magnetic fluxes under gauge groups $\tilde{a}$
  for which all $\br_{\tilde{a}}$ are proportional to each other, but
  not to at least one of the FI vectors of the remaining gauge groups
  $\br_{\hat{a}}$.
\end{itemize}
\end{theorem}
\emph{Proof}:
First let us prove that with all FI vectors parallel to each other,
the solution must be of type A.
Since $\bm\propto\sum_a\br_a k^a$ (see Eq.~\eqref{eq:1/2BPS_msol}), 
which are all proportional to each other, all $\br_a$ must be
orthogonal to $\bar\bell$, since $\bar\bell$ is orthogonal to $\bm$ by
definition and by Eq.~\eqref{eq:bell}.
If all $\br_a$ are orthogonal to $\bar\bell$,
i.e.~$\bar\bell\cdot\br_a=0$ for all $a$, then by proposition
\ref{prop:1}, the solution is of type A. 

Let us now prove that a single winding flavor always satisfies lemma
\ref{lemma:2}.
With just a single winding flavor $\tilde{A}\in S_{\rm WF}$,
$Q_{a\tilde{A}}^{-1}$ is just a vector in gauge group space,
$v_a\equiv Q_{a\tilde{A}}^{-1}$ and clearly
\beq
\sum_a\br_a v_a \propto\sum_a\br_a v_a n_{\tilde{A}},
\eeq
with $n_{\tilde{A}}\neq0$ being the winding number of the only winding
flavor.

Finally, let us prove the last statement of theorem \ref{thm:4}.
All the winding flavors $n_{\tilde{A}}\in S_{\rm WF}$ are charged
under a block in the charge matrix $Q$, which is block diagonal with
respect to the non-winding flavors $\hat{A}\notin S_{\rm WF}$:
\beq
Q_{Aa} =
\begin{pmatrix}
  \widetilde{Q}_{\tilde{A}\tilde{a}} & 0\\
  0 & \widehat{Q}_{\hat{A}\hat{a}}
\end{pmatrix},
\label{eq:Q_typeB2_blockdiagonal}
\eeq
and therefore the inverse of the charge matrix is given by
\beq
Q^{-1}_{aA} =
\begin{pmatrix}
  \widetilde{Q}^{-1}_{\tilde{a}\tilde{A}} & 0\\
  0 & \widehat{Q}^{-1}_{\hat{a}\hat{A}}
\end{pmatrix}.
\eeq
Due to the block-diagonal property of the inverse of the charge
matrix, the winding numbers turn on the following magnetic fluxes:
\beq
k^a = \sum_{\tilde{A}\in S_{\rm WF}}Q^{-1}_{a\tilde{A}} n_{\tilde{A}}
= \sum_{\tilde{A}\in S_{\rm WF}}\widetilde{Q}^{-1}_{\tilde{a}\tilde{A}} n_{\tilde{A}}
= k^{\tilde{a}},
\eeq
where we have used that only $n_{\tilde{A}}$ are nonvanishing. 
The ramification of this for the vector $\bm$ is
\beq
\bm \propto \sum_a \br_a k^a
= \sum_{\tilde{a}} \br_{\tilde{a}} k^{\tilde{a}}.
\eeq
Now using that all $\br_{\tilde{a}}$ are proportional to each other,
it follows that
\beq
\sum_a\br_aQ^{-1}_{a\tilde{A}}
= \sum_{\tilde{a}}\br_{\tilde{a}}\widetilde{Q}^{-1}_{\tilde{a}\tilde{A}},
\eeq
and
\beq
\sum_a\sum_{\tilde{B}\in S_{\rm WF}}\br_a Q^{-1}_{a\tilde{B}} n_{\tilde{B}}
= \sum_{\tilde{a}}\sum_{\tilde{B}\in S_{\rm WF}}\br_{\tilde{a}}
\widetilde{Q}^{-1}_{\tilde{a}\tilde{B}} n_{\tilde{B}}.
\eeq
Therefore, the two latter equations are proportional to each other as
required by lemma \ref{lemma:2} and the last statement of theorem
\ref{thm:4} follows.
\hfill$\square$

\begin{corollary}\label{coro:3}
A special case in which theorem \ref{thm:4} holds for type B2
solutions, is the case where the charge matrix $Q_{Aa}$ is diagonal. 
\end{corollary}

It is important that those $\psi_A\tilde\psi_A$ that do not vanish do
not wind and those that do wind do vanish.
In particular, it may seem nontrivial that the nonvanishing
$\psi_{\hat{A}}\tilde\psi_{\hat{A}}\neq 0$ satisfy
Eq.~\eqref{eq:constraint_nonwinding_flavors} with their vacuum
solution \eqref{eq:general_vac_sol}.
Intuitively, this clearly works out, because we have just changed the
basis of the constraint \eqref{eq:constraint_Phi_basis}, which is
indeed part of the vacuum equations. 

\begin{lemma}\label{lemma:3}
  For 1/2-BPS vortex solutions of type B, the nonwinding flavors obey
  the constraint \eqref{eq:constraint_nonwinding_flavors} by means of
  their vacuum solution \eqref{eq:general_vac_sol}.
\end{lemma}
\emph{Proof}:
Using the general vacuum solution \eqref{eq:general_vac_sol} and the
useful relations \eqref{eq:useful_vac_rels}, we can write
Eq.~\eqref{eq:constraint_nonwinding_flavors} as
\begin{align}
  2\langle\psi_{\hat{A}}\tilde\psi_{\hat{A}}\rangle &=
  \left(\frac12(1+m_3)(r_a^1-\i r_a^2)
  -\frac{m_-^2}{2(1+m_3)}(r_a^1+\i r_a^2)
  - m_-r_a^3\right) Q^{-1}_{a\hat{A}} \non
  &= \left(\frac{m_1m_3+\i m_2}{m_+}r_a^1
  +\frac{m_2m_3-\i m_1}{m_+}r_a^2
  -m_-r_a^3\right) Q^{-1}_{a\hat{A}},
\end{align}
and is automatically satisfied by the vacuum solution for all FI vectors
and all $\bm$, by using that $m_+m_-+m_3^2=1$.
\hfill$\square$

\begin{corollary}\label{coro:4}
  For 1/2-BPS vortex solutions of type B, the winding flavors
  satisfying Eq.~\eqref{eq:lemma2} of lemma \ref{lemma:2} will
  automatically have either $\psi_{\tilde{A}}\equiv 0$ or
  $\tilde\psi_{\tilde{A}}\equiv 0$.
\end{corollary}
\emph{Proof}:
By lemma \ref{lemma:2}, $\bOmega_{\tilde{A}}\propto\bm$ for all winding
flavors $\tilde{A}\in S_{\rm WF}$ and therefore by
Eq.~\eqref{eq:psiA_vac_mvec} either 
$|\langle\psi_{\tilde{A}}\rangle|^2=0$ or
$|\langle\tilde\psi_{\tilde{A}}\rangle|^2=0$.
Since the product $\psi_{\tilde{A}}\tilde\psi_{\tilde{A}}=0$ must
remain vanishing throughout $\mathbb{R}^2$ according to the constraint
\eqref{eq:constraint_psi2}, it follows that either
$\psi_{\tilde{A}}\equiv 0$ or $\tilde\psi_{\tilde{A}}\equiv 0$.
\hfill$\square$

The constraint equation \eqref{eq:constraint_psi2} is now satisfied
for all flavors. 
It remains to be proven that the would-be vortex equations are
satisfied by the inert flavors $\hat{A}$. 
This leads us to the next theorem.

\begin{theorem}\label{thm:5}
For non-winding flavors $\hat{A}$ of 1/2-BPS vortex solutions of type
B2 in the theory \eqref{eq:Lreduced}, the would-be vortex equations are
satisfied by the vacuum solution and moreover both the fields
$\psi_{\hat{A}}$ and $\tilde\psi_{\hat{A}}$ are individually constant
throughout $\mathbb{R}^2$. 
\end{theorem}
\emph{Proof}:
The field strength contracted with the charge matrix can be written as 
\beq
\sum_aQ_{Aa}F_{12}^a = 2\i\sum_a Q_{Aa} F_{\bar{z}z}^a
=2\p_{\bar{z}}\p_z\log|\psi_A|^2
=-2\p_{\bar{z}}\p_z\log|\tilde\psi_A|^2.
\label{eq:F12flavorspace}
\eeq
For winding flavors, the expression with the nonvanishing field
(i.e.~either $\psi_{\tilde{A}}$ or $\tilde\psi_{\tilde{A}}$) should be
used.
However, for the non-winding flavors $\hat{A}$ there are two cases:
If the left-hand side of Eq.~\eqref{eq:F12flavorspace} vanishes, both
$\psi_{\tilde{A}}$ and $\tilde\psi_{\tilde{A}}$ must be constant and
coincide with their respective VEVs.
If the left-hand-side of Eq.~\eqref{eq:F12flavorspace} does not
vanish, they are forced to be inversely proportional to each other. 

Now the right-hand side of the vortex equation \eqref{eq:BPS3_psi},
multiplied by the charge matrix, reads
\begin{align}
  &\sum_aQ_{Aa}e_a^2\bigg[
  \sum_B\left(|\psi_B|^2 - |\tilde\psi_B|^2\right) Q_{Ba}
  - \bm\cdot\br_a\bigg] \non
  &= \sum_aQ_{Aa}e_a^2\bigg[
  \sum_{\tilde{B}\in S_{\rm WF}}\left(|\psi_{\tilde{B}}|^2 - |\tilde\psi_{\tilde{B}}|^2\right) Q_{\tilde{B}a}
  +\sum_{\hat{B}\notin S_{\rm WF}}\left(|\psi_{\hat{B}}|^2 - |\tilde\psi_{\hat{B}}|^2\right) Q_{\hat{B}a}
  - \bm\cdot\br_a\bigg],
\end{align}
which we have split into winding flavors $\tilde{B}$ and non-winding
flavors $\hat{B}$ in the last line. 

Using now the block-diagonal property of the charge matrix
\eqref{eq:Q_typeB2_blockdiagonal} for the type B2 case, we obtain for
the non-winding flavors 
\beq
\sum_a Q_{\hat{A}a} F_{12}^a
= \sum_{\hat{a}} \widehat{Q}_{\hat{A}\hat{a}} F_{12}^{\hat{a}},
\eeq
for the left-hand side of the vortex equation \eqref{eq:BPS3_psi},
and
\begin{align}
  &\sum_aQ_{\hat{A}a}e_a^2\bigg[
  \sum_{\tilde{B}\in S_{\rm WF}}\left(|\psi_{\tilde{B}}|^2 - |\tilde\psi_{\tilde{B}}|^2\right) Q_{\tilde{B}a}
  +\sum_{\hat{B}\notin S_{\rm WF}}\left(|\psi_{\hat{B}}|^2 - |\tilde\psi_{\hat{B}}|^2\right) Q_{\hat{B}a}
  - \bm\cdot\br_a\bigg] \non
  &= \sum_{\hat{a}}\widehat{Q}_{\hat{A}\hat{a}}e_{\hat{a}}^2\bigg[
    \sum_{\hat{B}\notin S_{\rm WF}}\left(|\psi_{\hat{B}}|^2 - |\tilde\psi_{\hat{B}}|^2\right) \widehat{Q}_{\hat{B}\hat{a}}
    - \bm\cdot\br_{\hat{a}}\bigg].
\end{align}
This latter equation contains only the non-winding fields and thus
allows for the constant vacuum solution \eqref{eq:general_vac_sol}.
By insertion, we can verify that
\begin{align}
  &\sum_{\hat{a}}\widehat{Q}_{\hat{A}\hat{a}}e_{\hat{a}}^2\bigg[
    \sum_{\hat{B}\notin S_{\rm WF}}\left(|\langle\psi_{\hat{B}}\rangle|^2 - |\langle\tilde\psi_{\hat{B}}\rangle|^2\right) \widehat{Q}_{\hat{B}\hat{a}}
    - \bm\cdot\br_{\hat{a}}\bigg] \non
  &= \sum_{\hat{a}}\widehat{Q}_{\hat{A}\hat{a}}e_{\hat{a}}^2\bigg[
    \sum_{\hat{B}\notin S_{\rm WF}}\bm\cdot\bOmega_{\hat{B}}\widehat{Q}_{\hat{B}\hat{a}} -
    \bm\cdot\br_{\hat{a}}
    \bigg]\non
  &= \sum_{\hat{a}}\widehat{Q}_{\hat{A}\hat{a}}e_{\hat{a}}^2\left[
    \bm\cdot\br_{\hat{a}} - \bm\cdot\br_{\hat{a}}
    \right] = 0,
\end{align}
where we have used Eq.~\eqref{eq:psiA_vac_mvec} and that
\beq
\sum_{\hat{B}}\bOmega_{\hat{B}}\widehat{Q}_{\hat{B}\hat{a}}
=\sum_{\hat{B},\hat{b}}\br_{\hat{b}}\widehat{Q}^{-1}_{\hat{b}\hat{B}}\widehat{Q}_{\hat{B}\hat{a}}
=\br_{\hat{a}}.
\eeq
Since the vacuum solution \eqref{eq:psiA_vac_mvec} solves the
right-hand side of the vortex equation \eqref{eq:BPS3_psi} throughout
$\mathbb{R}^2$, the left-hand side vanishes as well and this is
consistent because the fluxes $k^{\hat{a}}$ are decoupled from the
winding flavors.
Therefore the constant solution is allowed and uniquely determined by
the vacuum equations.
\hfill$\square$

It should be clear from the latter proof that for type B1 solutions,
the would-be vortex equations do not decouple in general, and hence
they will induce nontrivial behavior in $\psi_{\hat{A}}$ and
$\tilde\psi_{\hat{A}}$ -- even though they are non-winding flavors --
but with
$\psi_{\hat{A}}\tilde\psi_{\hat{A}}=\langle\psi_{\hat{A}}\tilde\psi_{\hat{A}}\rangle$
everywhere constant and equal to their (product of) VEV(s). 

Analogously to the situation of type A solutions, we must have a
positive $n_{A}>0$ if $\langle\psi_A\rangle\neq 0$ and positive
$\tilde{n}_A=-n_A>0$ if $\langle\tilde\psi_A\rangle\neq 0$.
A slight complication arises because we can no longer ensure that the
FI vectors $\br_{\tilde{a}}$ are proportional to $\br_1$.
This, however, can quickly be remedied by a subsequent $\SU(2)_R$
rotation and a suitable relabeling of the gauge groups.

\begin{theorem}\label{thm:6}
For 1/2-BPS vortices of type B2 in the theory \eqref{eq:Lreduced}, the 
winding flavors $\tilde{A}$ turn on magnetic fluxes $k^{\tilde{a}}$
only in a subset of gauge groups $\tilde{a}$, whose FI vectors
$\br_{\tilde{a}}$ are all parallel by theorem \ref{thm:4} and
furthermore they can be chosen to be parallel with
$\br_1=(\alpha>0,0,0)$ without loss of generality. 
\end{theorem}
\emph{Proof}:
All the FI vectors $\tilde{a}$ are parallel to each other by theorem
\ref{thm:4} and can be rotated, without loss of generality to
$(\alpha'>0,0,0)$, for some positive constant $\alpha'$.
Now we will relabel the gauge groups so that one of them is $\tilde{a}=1$.
Dropping the prime $\alpha'\to\alpha$ completes the proof.
\hfill$\square$

\begin{corollary}\label{coro:5}
The vacuum of winding flavors $\tilde{A}$ in 1/2-BPS type B2 vortices
of the theory \eqref{eq:Lreduced} is given by the real solution
\beq
z_{\tilde{A}} =
\sum_{\tilde{a}}r_{\tilde{a}}^1\widetilde{Q}^{-1}_{\tilde{a}\tilde{A}},
\eeq
and the sign $m_1z_{\tilde{A}}$ determines whether the winding field
is $\psi_{\tilde{A}}$ (for $m_1z_{\tilde{A}}>0$) and hence
$n_{\tilde{A}}>0$ or $\tilde\psi_{\tilde{A}}$ (for
$m_1z_{\tilde{A}}<0$) and hence
$\tilde{n}_{\tilde{A}}=-n_{\tilde{A}}>0$.
\end{corollary}
\emph{Proof}:
Using the block-diagonal property of the charge matrix
\eqref{eq:Q_typeB2_blockdiagonal} for the type B2 solutions and
theorem \ref{thm:6}, the proof is analogous to that of corollary
\ref{coro:2}. 
\hfill$\square$

\begin{theorem}\label{thm:7}
  1/2-BPS solutions of type B2 in the theory \eqref{eq:Lreduced} must
  have magnetic fluxes
  \beq
  k^{\tilde{a}} = m_1\sum_{\tilde{A}\in S_{\rm WF},\tilde{b}}
  \widetilde{Q}^{-1}_{\tilde{a}\tilde{A}}c_{\tilde{A}}^2
  r_{\tilde{b}}^1\widetilde{Q}^{-1}_{\tilde{b}\tilde{A}}, \qquad
  c_{\tilde{A}}^2z_{\tilde{A}}\in\mathbb{Z},
  \eeq
  where the $c_{\tilde{A}}$ are adjusted so all winding numbers
  $|n_{\tilde{A}}|=c_{\tilde{A}}^2|z_{\tilde{A}}|$ are integers.
\end{theorem}
\emph{Proof}:
Using the block-diagonal property of the charge matrix
\eqref{eq:Q_typeB2_blockdiagonal} for the type B2 solutions and
corollary \ref{coro:5}, the proof is analogous to that of theorem
\ref{thm:2}.
\hfill$\square$

With theorem \ref{thm:7} in hand, we can finally rule out the
possibility of a solution with vanishing BPS bound.
\begin{theorem}\label{thm:8}
1/2-BPS solutions of type B2 in the theory \eqref{eq:Lreduced} cannot
have $\sum_ar_a^1k^a=0$, which would imply a vanishing energy bound
$T\geq T_{\rm BPS}=2\pi\left|\sum_a r_a^1k^a\right|=0$.
\end{theorem}
\emph{Proof}:
Using theorem \ref{thm:6} implying that
$\br_{\tilde{a}}=(r_{\tilde{a}}^1,0,0)$, the block-diagonal property
of the charge matrix \eqref{eq:Q_typeB2_blockdiagonal} implies via
theorem \ref{thm:7} that only $k^{\tilde{a}}$ are nonvanishing
(whereas $k^{\hat{a}}=0$). Finally, multiplying by $r_{\tilde{a}}^1$,
the proof is completed analogously to that of theorem \ref{thm:3}.
\hfill$\square$

For type B1 solutions, the single winding flavor turns on fluxes in
potentially all gauge groups, making it a complicated type of
solution.
We can nevertheless prove that it is impossible to have a vanishing
BPS bound also in this case.
\begin{theorem}\label{thm:9}
1/2-BPS solutions of type B1 in the theory \eqref{eq:Lreduced} cannot
have $\sum_a\br_a k^a=0$, which would imply a vanishing energy bound
$T\geq T_{\rm BPS}=2\pi\sqrt{\sum_{a,b}\br_a\cdot\br_bk^a k^b}=0$.
\end{theorem}
\emph{Proof}:
By the definition of the type B1 solution according to theorem
\ref{thm:4}, only a single flavor possesses nonvanishing winding number
$n_{\tilde{A}}\neq 0$ and therefore the magnetic fluxes are given by
\beq
k^a = Q^{-1}_{a\tilde{A}} n_{\tilde{A}},
\eeq
where there is no sum over $\tilde{A}$ since it is just a single
flavor.
We can thus write
\beq
\sum_a\br_ak^a = \sum_a\br_a Q^{-1}_{a\tilde{A}}n_{\tilde{A}}
= \bOmega_{\tilde{A}} n_{\tilde{A}} \neq 0,
\eeq
where we have used Eq.~\eqref{eq:Omega_def} and the fact
that neither $n_{\tilde{A}}$ can vanish nor can $\bOmega_{\tilde{A}}$.
The latter follows from Eq.~\eqref{eq:bOmega_cannot_vanish}, which
states that $|\bOmega_{\tilde{A}}|=0$ corresponds to
$|\langle\psi_{\tilde{A}}\rangle|^2+|\langle\tilde\psi_{\tilde{A}}\rangle|^2=0$, 
which in turn implies unbroken gauge symmetry which we cannot allow.
\hfill$\square$

In the type B1 solution, it remains whether there is any restriction
on the sign of the (only nonvanishing) winding number $n_{\tilde{A}}$
with regards to the vacuum solution.
This leads us to the following lemma.

\begin{lemma}\label{lemma:4}
The sign of the only nonvanishing winding number $n_{\tilde{A}}\neq0$,
in 1/2-BPS solutions of type B1 in the theory \eqref{eq:Lreduced}, is
not restricted by the vacuum solution.
\end{lemma}
\emph{Proof}:
The unit vector $\bm$ is directed as
\beq
\bm\propto\sum_a\br_ak^a=\sum_a\br_aQ^{-1}_{a\tilde{A}}n_{\tilde{A}},
\eeq
and the sign of $n_{\tilde{A}}$ thus determines whether $\bm$ is
parallel ($n_{\tilde{A}}>0$) or anti-parallel ($n_{\tilde{A}}<0$) with
$\bOmega_{\tilde{A}}$ as
\beq
\bOmega_{\tilde{A}}\equiv \sum_a\br_a Q^{-1}_{a\tilde{A}}.
\eeq
Thus, if they are parallel ($n_{\tilde{A}}>0$) then by
Eq.~\eqref{eq:psiA_vac_mvec}, $\tilde\psi_{\tilde{A}}\equiv0$ and
\emph{vice versa} if they are anti-parallel ($n_{\tilde{A}}<0$) then
$\psi_{\tilde{A}}\equiv0$.
Since, there is only a single winding field, no incompatibility can
arise and the lemma follows.
\hfill$\square$

\subsection{Master equations}

For supersymmetric (BPS) solitons, it is often the case that there is a
selfdual equation of the form $D_{\bar{z}}\Phi=0$ and another BPS
equation of the form $F_{12}=\cdots$, which generically can be
combined into a resulting gauge-invariant equation in terms of a
non-holomorphic field, independent of the number of codimensions
(i.e.~whether the system contains domain walls, vortices, etc.) or of
the gauge symmetry (Abelian or non-Abelian) or other specifics of the
system, see Refs.~\cite{Isozumi:2004vg,Eto:2005yh,Eto:2006pg}.
In our case, that field is $\varsigma_A(z,\bar{z})$ and the master
equation is the governing equation for that field. 

We will now write down the master equations for the different types
of solution in turn, starting with type A.

\subsubsection{Type A solutions}

By proposition \ref{prop:1}, type A solutions solve the constraint
equation \eqref{eq:constraint_psi} by having $\psi_A\tilde\psi_A=0$
for all flavors throughout $\mathbb{R}^2$, which means either $\psi_A$
or $\tilde\psi_A$ can be a winding field and the other vanishes
everywhere.
By theorem \ref{thm:1}, $\br_a\propto\br_1$ for all gauge groups $a$
and thus without loss of generality $\br_a=(r_a^1,0,0)$ and in turn
$\bm=(\pm1,0,0)$.
The sign $m_1$ is given by the sign of $\sum_ar_a^1k^a$, which cannot
have vanishing length by theorem \ref{thm:3} and the magnetic fluxes
take the form
\beq
k^a = m_1\sum_{A,b}Q^{-1}_{aA}c_A^2r_b^1Q^{-1}_{bA}, \qquad
c_A^2z_A\in\mathbb{Z},
\eeq
where $c_A\in\mathbb{R}$ are constants adjusted so the winding numbers
$|n_A|=c_A^2|z_A|$ are all integers, by theorem \ref{thm:2}.
Therefore the BPS tension is always nonvanishing for all existing
solutions of type A and the vortex equations can be written in the
form of the following master equations.

\begin{theorem}\label{thm:10}
1/2-BPS solutions of type A in the theory \eqref{eq:Lreduced}, with
the above described properties, are governed by the master equations
\begin{align}
  -2\sign(z_A)\p_{\bar{z}}\p_z\log|\varsigma_A(z,\bar{z})|^2
  = \sum_{a,B} Q_{Aa} e_a^2 \big(Q^{\rm T}\big)_{aB} z_B
      \left(\left|\frac{h_B(z)}{\varsigma_B(z,\bar{z})}\right|^2-1\right),
      \label{eq:thm10}
\end{align}
with $z_A=\sum_ar_a^1Q^{-1}_{aA}$ and $h_B(z)$ a holomorphic
polynomial of degree $|n_B|$ of the form
\beq
h_B(z)=\prod_{i=1}^{|n_B|}(z-Z_i^B),
\eeq
with $Z_i^B\in\mathbb{C}$ being position moduli and if $n_B=0$, $h_B=1$
can be chosen without loss of generality.
Finally, the boundary condition on $\varsigma_A(z,\bar{z})$ is given
by $\lim_{|z|\to\infty}|\varsigma_A(z,\bar{z})|=|z|^{2|n_A|}$. 
\end{theorem}
\emph{Proof}:
We start by multiplying both sides of the vortex equation
\eqref{eq:BPS3_psi} by $Q_{Aa}$ (and sum over the gauge index $a$) 
\beq
\sum_a Q_{Aa}F_{12}^a
= \sum_a Q_{Aa} e_a^2\bigg(
\sum_B Q_{aB}^{\rm T}\left(|\psi_B|^2 - |\tilde\psi_B|^2\right)
- m_1 r_a^1\bigg),
\label{eq:BPS3_flavor}
\eeq
where we have used that only $r_a^1$ is nonvanishing (of the FI
vector components).
For each flavor we will switch variables into the moduli functions
$h_A(z)$ and auxiliary fields $\varsigma_A(z,\bar{z})$, according to
which field is the winding field:
\beq
\psi_A = \sqrt{|z_A|}\frac{h_A(z)}{\varsigma_A(z,\bar{z})}, \qquad\textrm{or}\qquad
\tilde\psi_A = \sqrt{|z_A|}\frac{h_A(z)}{\varsigma_A(z,\bar{z})}.
\eeq
That in turn is decided by the sign of
$m_1z_A=m_1\sum_ar_a^1Q^{-1}_{aA}$, which when nonvanishing has the
same sign as $n_A$ according to corollary \ref{coro:2} and we have
multiplied by the norm of the field's VEV $\sqrt{|z_A|}$.
According to lemma \ref{lemma:1} and proposition \ref{prop:1}, this
sign determines whether $\psi_A$ or $\tilde\psi_A$ is the nonvanishing
flavor: for $n_A>0$ we have $\tilde\psi_A\equiv0$ and $\psi_A$ is the
winding fields and \emph{vice versa} for negative $n_A$.
When $n_A=0$, the nonvanishing field is still determined by the sign
of $m_1z_A$ and its would-be vortex equation is the same master
equation as that of a winding flavor.
We therefore have
\begin{align}
  -2\sign(m_1z_A)\p_{\bar{z}}\p_z\log|\varsigma_A(z,\bar{z})|^2
  =\sum_{a,B}Q_{Aa}e_a^2\big(Q^{\rm T}\big)_{aB}\sign(m_1z_B)|z_B|
  \left|\frac{h_B(z)}{\varsigma(z,\bar{z})}\right|^2
  -\sum_a Q_{Aa}e_a^2m_1r_a^1.
\end{align}
Now, since $\sign(AB)=\sign(A)\sign(B)$ and $\sign(m_1)=m_1=\pm1$,
$m_1$ drops out of the entire master equation, $\sign(z_B)|z_B|=z_B$,
and only the sign of the vacuum solution $z_A$ determines the sign of
the kinetic term.
The constant (last) term can be rewritten as follows
\beq
-\sum_aQ_{Aa}e_a^2r_a^1
=-\sum_{a,b,B}Q_{Aa}e_a^2 Q_{Ba} r_b^1 Q_{Bb}^{-1} 
=-\sum_{a,b,B}Q_{Aa}e_a^2 \big(Q^{\rm T}\big)_{aB} z_B.
\eeq
We are thus left with Eq.~\eqref{eq:thm10}.
Finally, the degree of the polynomial $h_A(z)$ for $n_A>0$ must be
$n_A$, however, since we have given the same name to the
moduli function of the anti-fundamental field, the degree in the case
that $\tilde\psi_A$ is nonvanishing must have $h_A(z)$ of degree
$\tilde{n}_A=-n_A$.
In all cases, the degree of the polynomial is thus given by $|n_A|$
and the theorem follows.
\hfill$\square$

\subsubsection{Type B2 solutions}

By proposition \ref{prop:1}, the type B solutions are characterized by
having $\bar\bell\cdot\br_a\neq0$ for some gauge group indices $a$ and
therefore not all FI vectors are parallel.
For convenience, we split the flavors into winding flavors
$\tilde{A}\in S_{\rm WF}$ and non-winding flavors
$\hat{A}\notin S_{\rm WF}$. 
For type B2 solutions according to theorem \ref{thm:4} and by lemma
\ref{lemma:2} all winding flavors $\tilde{A}$ are via the charge
matrix coupled only to parallel FI vectors $\tilde{a}$ due to the
block-diagonal property \eqref{eq:Q_typeB2_blockdiagonal} of the
charge matrix and by theorem \ref{thm:6}, those FI vectors can all be
taken to be proportional to $\br_1=(\alpha>0,0,0)$ without loss of
generality. 
The block-diagonal property \eqref{eq:Q_typeB2_blockdiagonal} of the
charge matrix further implies that magnetic fluxes are only turned on
in gauge groups with parallel FI vectors and therefore
$\bm=(\pm1,0,0)$.
Furthermore, the non-winding flavors satisfy -- by means of their
vacuum solution -- the constraint equations according to lemma
\ref{lemma:3} and the would-be vortex equations according to theorem
\ref{thm:5}. 
For the winding flavors $\tilde{A}$, either $\psi_{\tilde{A}}\equiv0$
or $\tilde\psi_{\tilde{A}}\equiv0$ by corollary \ref{coro:4} and lemma
\ref{lemma:2} and by corollary \ref{coro:5} if $m_1z_{\tilde{A}}>0$
then $\tilde\psi_{\tilde{A}}\equiv0$ and $\psi_{\tilde{A}}$ is the
winding field with winding number $n_{\tilde{A}}\geq 0$, whereas if
$m_1z_{\tilde{A}}<0$ then $\psi_{\tilde{A}}\equiv0$ and
$\tilde\psi_{\tilde{A}}$ is the winding field with winding number
$\tilde{n}_{\tilde{A}}=-n_{\tilde{A}}\geq0$.
By theorem \ref{thm:7} the magnetic fluxes are given by
\beq
k^{\tilde{a}} = m_1\sum_{\tilde{A}\in S_{\rm WF},\tilde{b}}
\widetilde{Q}^{-1}_{\tilde{a}\tilde{A}}c_{\tilde{A}}^2r_{\tilde{b}}^1
\widetilde{Q}^{-1}_{\tilde{b}\tilde{A}}, \qquad
c_{\tilde{A}}^2z_{\tilde{A}}\in\mathbb{Z},
\eeq
where the $c_{\tilde{A}}$ are adjusted so all winding numbers
$|n_{\tilde{A}}|=c_{\tilde{A}}^2|z_{\tilde{A}}|$ are integers and
$m_1$ is given by the sign of
$\sum_{\tilde{a}}r_{\tilde{a}}^1k^{\tilde{a}}$, which in turn cannot
vanish by theorem \ref{thm:8}.
Therefore the BPS tension is always nonvanishing for all existing
solutions of type B2 and the vortex equations can be written in the
form of the following master equations.

\begin{theorem}\label{thm:11}
All winding flavors $\tilde{A}$ in 1/2-BPS solutions of type B2 in the
theory \eqref{eq:Lreduced}, with the above described properties, are
governed by the master equations
\begin{align}
  -2\sign(z_{\tilde{A}})\p_{\bar{z}}\p_z\log|\varsigma_{\tilde{A}}(z,\bar{z})|^2
  = \sum_{\tilde{a},\tilde{B}} \widetilde{Q}_{\tilde{A}\tilde{a}}
  e_{\tilde{a}}^2 \big(\widetilde{Q}^{\rm T}\big)_{\tilde{a}\tilde{B}}z_{\tilde{B}}
      \left(\left|\frac{h_{\tilde{B}}(z)}{\varsigma_{\tilde{B}}(z,\bar{z})}\right|^2-1\right),
\end{align}
with
$z_{\tilde{A}}=\sum_{\tilde{a}}r_{\tilde{a}}^1\widetilde{Q}^{-1}_{\tilde{a}\tilde{A}}$
and $h_{\tilde{B}}(z)$ a holomorphic polynomial of degree
$|n_{\tilde{B}}|$ of the form 
\beq
h_{\tilde{B}}(z)=\prod_{i=1}^{|n_{\tilde{B}}|}(z-Z_i^B),
\eeq
with $Z_i^B\in\mathbb{C}$ being position moduli and if
$n_{\tilde{B}}=0$ (which we shall allow to cover all solution types),
$h_{\tilde{B}}=1$ can be chosen without loss of generality.
Finally, the boundary condition on $\varsigma_{\tilde{A}}(z,\bar{z})$
is given by
$\lim_{|z|\to\infty}|\varsigma_{\tilde{A}}(z,\bar{z})|=|z|^{2|n_{\tilde{A}}|}$. 
\end{theorem}
\emph{Proof}:
Using the block-diagonal property \eqref{eq:Q_typeB2_blockdiagonal} of 
the charge matrix and the above mentioned properties of type B2
solutions, the proof is analogous to that of theorem \ref{thm:10}.
\hfill$\square$

\subsubsection{Type B1 solutions}

By proposition \ref{prop:1}, the type B solutions have
$\bar\bell\cdot\br_a\neq0$ for some gauge group indices $a$ and
therefore the FI vectors are not all parallel.
The type B1 solution, by theorem \ref{thm:4} is defined by having only
a single winding flavor $\tilde{A}$ for which we must have
$n_{\tilde{A}}\neq 0$. The unit vector $\bm\propto\sum_a\br_ak^a$ and
the latter cannot be vanishing by theorem \ref{thm:9}. 
Either $\psi_{\tilde{A}}$ is winding (for $n_{\tilde{A}}>0$) and
$\tilde\psi_{\tilde{A}}\equiv0$ or $\tilde\psi_{\tilde{A}}$ is winding
(for $\tilde{n}_{\tilde{A}}=-n_{\tilde{A}}>0$) and
$\psi_{\tilde{A}}\equiv0$.
By lemma \ref{lemma:4} there is no restriction on the sign of
$n_{\tilde{A}}$.
Since there is no restriction on the charge matrix (other than
$\det Q\neq0$), the master equations change to the following form.

\begin{theorem}\label{thm:12}
1/2-BPS solutions of type B1 in the theory \eqref{eq:Lreduced}, with
the above described properties, are governed by the master equations
\begin{align}
  -2\Upsilon_A\p_{\bar{z}}\p_z\log|\varsigma_A(z,\bar{z})|^2
  &=\sum_{a}Q_{Aa}e_a^2\big(Q^{\rm T}\big)_{a\tilde{A}}
  \sign(n_{\tilde{A}})|\bOmega_{\tilde{A}}|
  \left(\left|\frac{h_{\tilde{A}}(z)}{\varsigma_{\tilde{A}}(z,\bar{z})}\right|^2-1\right)\label{eq:thm12}\\
  &\phantom{=\ }+\frac12\sum_{a,\hat{B}\notin S_{\rm WF}}Q_{Aa}e_a^2\big(Q^{\rm T}\big)_{a\hat{B}}
  \left(|\bOmega_{\hat{B}}| + \bm\cdot\bOmega_{\hat{B}}\right)
  \left(\frac{1}{|\varsigma_{\hat{B}}(z,\bar{z})|^2}-1\right)\non
  &\phantom{=\ }-\frac12\sum_{a,\hat{B}\notin S_{\rm WF}}Q_{Aa}e_a^2\big(Q^{\rm T}\big)_{a\hat{B}}
  \left(|\bOmega_{\hat{B}}| - \bm\cdot\bOmega_{\hat{B}}\right)
  \left(|\varsigma_{\hat{B}}(z,\bar{z})|^2-1\right),\nonumber
\end{align}
with
\beq
\Upsilon_A=
\begin{cases}
  \sign(n_{\tilde{A}}), & A=\tilde{A},\\
   1, & \textrm{otherwise},
\end{cases}\qquad
\bOmega_A=\sum_a\br_aQ^{-1}_{aA}, \qquad
\bm=\frac{\sum_a\br_ak^a}{\sqrt{\sum_{b,c}\br_b\cdot\br_ck^bk^c}},
\eeq
and $h_{\tilde{A}}(z)$ a holomorphic polynomial of degree
$|n_{\tilde{A}}|$ of the form
\beq
h_{\tilde{A}}(z) = \prod_{i=1}^{|n_{\tilde{A}}|}(z - Z_i),
\eeq
with $Z_i\in\mathbb{C}$ being position moduli.
Finally, the boundary condition for the (only) winding auxiliary field
$\varsigma_{\tilde{A}}(z,\bar{z})$ is
$\lim_{|z|\to\infty}\varsigma_{\tilde{A}}(z,\bar{z})=|z|^{2|n_{\tilde{A}}|}$ and
for the non-winding auxiliary fields $\varsigma_{\hat{A}}(z,\bar{z})$ it is
$\lim_{|z|\to\infty}\varsigma_{\hat{A}}(z,\bar{z})=1$. 
\end{theorem}
\emph{Proof}:
Using the vortex equation \eqref{eq:BPS3_psi} multiplied by $Q_{Aa}$
(and summed over $a$), we have Eq.~\eqref{eq:BPS3_flavor}.
Now for the winding flavor $\tilde{A}$, we will switch variables into
the moduli function $h_{\tilde{A}}(z)$ and the auxiliary field
$\varsigma_{\tilde{A}}(z,\bar{z})$, according to which fields is the
winding field:
\beq
\psi_{\tilde{A}} = \sqrt{|\bOmega_{\tilde{A}}|}
\frac{h_{\tilde{A}}(z)}{\varsigma_{\tilde{A}}(z,\bar{z})}, \qquad\textrm{or}\qquad
\tilde\psi_{\tilde{A}} = \sqrt{|\bOmega_{\tilde{A}}|}\frac{h_{\tilde{A}}(z)}{\varsigma_{\tilde{A}}(z,\bar{z})},
\label{eq:winding_fields_B1}
\eeq
where we have used that the vacuum of the nonvanishing (and hence
winding) field is $\sqrt{|\bOmega_{\tilde{A}}|}$ and which field is
winding is determined by $\sign(n_{\tilde{A}})$.
We thus have
\beq
\sum_a Q_{Aa}F_{12}^a
=-2\sign(n_{\tilde{A}})\p_{\bar{z}}\p_z\log|\varsigma_{\tilde{A}}(z,\bar{z})|^2.
\eeq
Solving the self-dual equations
$D_{\bar{z}}\psi_A=D_{\bar{z}}\tilde\psi_A=0$, works by picking the
nonvanishing field and determining $A_{\bar{z}}$ from that field
($\psi_A$ or $\tilde\psi_A$) and it is consistent, because the other
field being everywhere zero solves the other self-dual equation.
For the non-winding flavors, the product
$\psi_{\hat{A}}\tilde\psi_{\hat{A}}$ may not vanish, but is everywhere
constant and equal to its VEV
$\langle\psi_{\hat{A}}\tilde\psi_{\hat{A}}\rangle$, according to lemma
\ref{lemma:3}.
Therefore, in order to solve both self-dual equations, we set
\beq
\psi_{\hat{A}}=\frac{\sqrt{|\bOmega_{\hat{A}}|+\bm\cdot\bOmega_{\hat{A}}}}{\sqrt{2}\varsigma_{\hat{A}}(z,\bar{z})},\qquad
\tilde\psi_{\hat{A}}=\sqrt{|\bOmega_{\hat{A}}|-\bm\cdot\bOmega_{\hat{A}}}\frac{\varsigma_{\hat{A}}(z,\bar{z})}{\sqrt{2}},
\label{eq:nonwinding_ansatz_B1}
\eeq
which is consistent with
\beq
\sum_a Q_{Aa}F_{12}^a
=2\p_{\bar{z}}\p_z\log|\psi_{\hat{A}}|^2
=-2\p_{\bar{z}}\p_z\log|\tilde\psi_{\hat{A}}|^2
=-2\p_{\bar{z}}\p_z\log|\varsigma_{\hat{A}}(z,\bar{z})|^2,
\eeq
and the sign is fixed by the Ansatz (definition)
\eqref{eq:nonwinding_ansatz_B1}.
Inserting the fields of Eqs.~\eqref{eq:winding_fields_B1} and
\eqref{eq:nonwinding_ansatz_B1} into the right-hand side of
Eq.~\eqref{eq:BPS3_flavor}, we obtain the result \eqref{eq:thm12}
apart from the constant term, which needs a bit more massage
\begin{align}
-\sum_a Q_{Aa}e_a^2\bm\cdot\br_a
&=-\sum_{a,B,b} Q_{Aa}e_a^2\big(Q^{\rm T}\big)_{aB} Q^{-1}_{bB}\bm\cdot\br_b \non
&=-\sum_{a,B} Q_{Aa}e_a^2\big(Q^{\rm T}\big)_{aB} \bm\cdot\bOmega_B \non
&=-\sum_{a} Q_{Aa}e_a^2\big(Q^{\rm T}\big)_{a\tilde{A}} \bm\cdot\bOmega_{\tilde{A}}
-\sum_{a,\hat{B}\notin S_{\rm WF}} Q_{Aa}e_a^2\big(Q^{\rm T}\big)_{a\hat{B}}
\bm\cdot\bOmega_{\hat{B}} \non
&=-\sum_{a} Q_{Aa}e_a^2\big(Q^{\rm T}\big)_{a\tilde{A}} \sign(n_{\tilde{A}})|\bOmega_{\tilde{A}}|
-\sum_{a,\hat{B}\notin S_{\rm WF}} Q_{Aa}e_a^2\big(Q^{\rm T}\big)_{a\hat{B}}
\bm\cdot\bOmega_{\hat{B}},
\end{align}
where we have used the definition \eqref{eq:Omega_def}, split the sum
over the flavor index $B$ into the (single) winding index $\tilde{A}$
and the non-winding flavors $\hat{B}$, and finally used that
$\bOmega_{\tilde{A}}$ must be parallel or anti-parallel with $\bm$
(which is a unit vector) by corollary \ref{coro:4} and
Eq.~\eqref{eq:psiA_vac_mvec}; the sign of whether the vectors are
parallel or anti-parallel is that of $n_{\tilde{A}}$ by definition. 
We thus arrive at the result in Eq.~\eqref{eq:thm12}. 
The proof of the moduli function is analogous to that of theorem
\ref{thm:10}.
\hfill$\square$

\subsection{Vortex equations in standard form}\label{sec:Taubes_eqs}

It will prove convenient to rewrite the master equations of the last
subsection to a more commonly used form, i.e.~like the Taubes
equation.

\subsubsection{Type A solutions}

Starting with type A solutions, we make the change of variables
$u_A\equiv-2\log\left|\frac{h_A(z)}{\varsigma_A(z,\bar{z})}\right|$,
and obtain from theorem \ref{thm:10},
\beq
\nabla^2u_A = \sum_B\mathcal{A}_{AB}
\left(e^{u_B}-1\right) + 4\pi\sum_{i=1}^{|n_A|}\delta(z - Z_i^A),
\label{eq:Taubes_A}
\eeq
where we have defined the matrix
\beq
\mathcal{A}_{AB} \equiv
2\sum_a Q_{Aa}e_a^2Q_{Ba} \sign(z_A)z_B,
\eeq
which is real and positive definite, but not symmetric.
The boundary conditions on $u_A$ are $\lim_{|z|\to\infty}u_A=0$,
$\forall A$. 

In the special case where the matrix $\mathcal{A}$ is symmetric, the
existence and uniqueness (as well as sharp decay estimates) have been
proven by Yang in Ref.~\cite{Yang:2000in}.
The matrix can be made symmetric, by setting the same size for all the
VEVs as
\beq
z_B = \pm \alpha,
\eeq
with a sign that can be chosen arbitrarily for each flavor.

In the general case, where the matrix is only positive definite, but
not symmetric, we can use the proof of existence in Yang's book
\cite[Chapter~4.7]{Yang:2004}.
In order to use this existence proof, we must have that
\beq
\sum_A \mathcal{A}^{-1}_{BA} g_A > 0,
\label{eq:general_cond}
\eeq
for the equation
\beq
\nabla^2u_A = \sum_B \mathcal{A}_{AB} e^{u_B} - g_A.
\eeq
Since $\mathcal{A}$ is invertible (it is not possessing any vanishing
eigenvalues), we can readily calculate the condition
\eqref{eq:general_cond} for our case:
\beq
\sum_{A,C} \big(\mathcal{A}^{-1}\big)_{BA} \mathcal{A}_{AC}
=\sum_C \delta_{BC} = 1 > 0, \qquad \forall B,
\eeq
which indeed is positive.
We do not know of any uniqueness results in this case. 

As mentioned in the introduction, the vortex equations found here are
identical to those of Ref.~\cite{Schroers:1996zy}, which is a
similar model, but with only $\mathcal{N}=1$ supersymmetry and hence
no $\SU(2)_R$ symmetry. 
Therefore, solving the constraint equations in the $\mathcal{N}=2$
theory basically reduce the vortex equations to those of an
$\mathcal{N}=1$ theory.

\subsubsection{Type B2 solutions}

Clearly, the same equation is obtained for type B2 solutions, from
theorem \ref{thm:11}, for the winding flavors 
\beq
\nabla^2u_{\tilde{A}} = \sum_{\tilde{B}}\mathcal{A}_{\tilde{A}\tilde{B}}
\left(e^{u_{\tilde{B}}}-1\right) + 4\pi\sum_{i=1}^{|n_{\tilde{A}}|}\delta(z - Z_i^{\tilde{A}}),
\label{eq:Taubes_B2}
\eeq
and the matrix is now
\beq
\mathcal{A}_{\tilde{A}\tilde{B}} \equiv
2\sum_{\tilde{a}} \widetilde{Q}_{\tilde{A}\tilde{a}}e_{\tilde{a}}^2
\widetilde{Q}_{\tilde{B}\tilde{a}} \sign(z_{\tilde{A}})z_{\tilde{B}},
\eeq
which is still real and positive definite, but not symmetric.
Clearly the same results of existence and uniqueness applies to this
case as to the case of type A solutions, see the previous
subsubsection.
The boundary conditions on $u_{\tilde{A}}$ are $\lim_{|z|\to\infty}u_{\tilde{A}}=0$,
$\forall {\tilde{A}}$.

\subsubsection{Type B1 solutions}

Finally, for the type B1 solutions a new type of vortex equation is
found in theorem \ref{thm:12}, and in standard form it reads
\begin{equation}
\nabla^2u_A
= \mathcal{A}_A\left(e^{u_N}-1\right)
+ \sum_{B=1}^{N-1}\mathcal{B}_{AB}\left(e^{u_B}-1\right)
- \sum_{B=1}^{N-1}\mathcal{C}_{AB}\left(e^{-u_B}-1\right)
+4\pi\delta^{AN}\sum_{i=1}^{|n_N|}\delta(z-Z_i),
\label{eq:Taubes_B1}
\end{equation}
where we have selected the last flavor as the winding flavor and
defined the vector
\beq
\mathcal{A}_A\equiv
2\sum_aQ_{Aa}e_a^2Q_{Na}\sign(n_N)|\bOmega_N|,
\eeq
as well as the $N$-by-$(N-1)$ matrices
\begin{align}
  \mathcal{B}_{AB}&\equiv
  \sum_a\sum_{B\neq N} Q_{Aa}e_a^2Q_{Ba}\left(|\bOmega_B|+\bm\cdot\bOmega_B\right),\\
  \mathcal{C}_{AB}&\equiv
  -\sum_a\sum_{B\neq N} Q_{Aa}e_a^2Q_{Ba}\left(|\bOmega_B|-\bm\cdot\bOmega_B\right).
\end{align}
The boundary conditions on $u_A$ are $\lim_{|z|\to\infty}u_A=0$,
$\forall A$ as well as for $A=N$. 
We do not know of any existence or uniqueness results for this case,
which is to the best of our knowledge a new type of vortex equation.

\subsection{Explicit examples}\label{sec:examples}

We will now flesh out explicit examples of 1/2-BPS solutions to the
constraint equation \eqref{eq:constraint_psi2} and display the
corresponding master equations obtained by theorems
\ref{thm:10}-\ref{thm:12} as well as vortex equations in the standard
form, found in the previous subsection. 

\subsubsection{\texorpdfstring{$N=M=1$}{N=M=1}}\label{sec:example_N=M=1}

For one gauge group ($N=1$) and one flavor ($M=1$), the indices are
just cluttering, so we will suppress them here.
The vacuum solution is given by Eq.~\eqref{eq:vacsol_N=M=1}, the FI
vector is $\br=(\alpha>0,0,0)$, and the solution is always of type A
and either fully fundamental or fully anti-fundamental\footnote{We do
  not allow $n=0$, $Q=0$ and therefore neither $k=0$, since that is
  the (trivial) vacuum solution.}, with string tension
\beq
T_{\rm BPS} = 2\pi|k|\alpha.
\eeq
The master equation is thus given by theorem \ref{thm:10} and reads
\beq
-2\p_{\bar{z}}\p_z\log|\varsigma(z,\bar{z})|^2
= |Q|e^2\alpha\left(\left|\frac{h(z)}{\varsigma(z,\bar{z})}\right|^2
- 1\right),
\label{eq:master_N=M=1}
\eeq
which holds for any combination of signs of $Q$ and $k$ (and in
turn of $n=Qk$).
It will prove instructive to see explicitly how the constraint
equations are solved in this simple situation. 
The $\SU(2)_R$ vectors read
\beq
\bm = (\sign(k),0,0),\qquad
\bar\bell = -\frac{1}{\sqrt{2}} (0,\i,\sign(k)),
\eeq
by theorem \ref{thm:1} so it is manifest that $\bar\bell$ and $\br$
are orthogonal; on the 
other hand it is obvious that the constraint
\eqref{eq:typeA_constraint2} is always satisfied. 
The constraint \eqref{eq:typeA_constraint1} is more nontrivial and
becomes 
\begin{align}
  \sign(n) =
  \begin{cases}
  -1, &\quad\Rightarrow\qquad
  \psi(z,\bar{z}) \equiv 0, \qquad
  \tilde\psi(z,\bar{z}) = \varsigma^{-1}(z,\bar{z})h(z), \\
  +1, &\quad\Rightarrow\qquad
  \tilde\psi(z,\bar{z}) \equiv 0, \qquad
  \psi(z,\bar{z}) = \varsigma^{-1}(z,\bar{z})h(z).
  \end{cases}
\end{align}
By rescaling the lengths $z\to \sqrt{2|Q|\alpha}e z$ and defining a
field $u\equiv2\log\left|\frac{h(z)}{\varsigma(z,\bar{z})}\right|$, we can write
the master equation in the form
\beq
\nabla^2u = e^u - 1 + 4\pi\sum_{i=1}^{|n|}\delta(z-Z_i),
\eeq
which is Taubes equation (Eq.~\eqref{eq:Taubes_A}).
The existence and uniqueness has been proved by Taubes in
Ref.~\cite{Taubes:1979tm}.

\subsubsection{\texorpdfstring{$N=M=2$}{N=M=2}}

We now turn to the case of two gauge groups with two flavors
($N=M=2$) in which case there are solutions of all types.

\paragraph[Type A solution]{Type A solution.}\label{sec:example_N=M=2_type}

By theorem \ref{thm:1} the FI vectors are parallel and hence read
$\br_1=(\alpha,0,0)$ and $\br_2=(\beta,0,0)$ with $\alpha>0$ and
$\beta\neq 0$ and in turn $\bm=(\pm1,0,0)$ (by theorem
\ref{thm:1}). The vacuum solution is given by
$z_A=\sum_ar_a^1Q^{-1}_{aA}$ (corollary \ref{coro:2}) and the string
tension reads
\beq
T_{\rm BPS} = 2\pi|\alpha k^1 + \beta k^2|>0.
\eeq
We will now select an explicit example for the charge matrix with
nonvanishing determinant
\begin{align}
Q =
\begin{pmatrix}
  1 & q\\
  0 & 1
\end{pmatrix}, \qquad
q\in\mathbb{Z},
\end{align}
for which the vacuum solution is given by
\beq
z_A = (\alpha, \beta - q\alpha),
\eeq
and the master equations by theorem \ref{thm:10} read
\begin{align}
-2\p_{\bar{z}}\p_z\log|\varsigma_1(z,\bar{z})|^2
&= (e_1^2+q^2e_2^2)\alpha\left(\left|\frac{h_1(z)}{\varsigma_1(z,\bar{z})}\right|^2-1\right)
+qe_2^2(\beta-q\alpha)\left(\left|\frac{h_2(z)}{\varsigma_2(z,\bar{z})}\right|^2-1\right),\label{eq:master_N=M=2_1}\\
-2\p_{\bar{z}}\p_z\log|\varsigma_2(z,\bar{z})|^2
&= \sign(\beta-q\alpha)qe_2^2\alpha\left(\left|\frac{h_1(z)}{\varsigma_1(z,\bar{z})}\right|^2-1\right)
+e_2^2|\beta-q\alpha|\left(\left|\frac{h_2(z)}{\varsigma_2(z,\bar{z})}\right|^2-1\right).\label{eq:master_N=M=2_2}
\end{align}
Note that $\beta-q\alpha\neq0$ cannot vanish as that would make the
theory touch the Coulomb branch on the second flavor, or in other
words, the gauge symmetry would be unbroken.

By theorem \ref{thm:2}, the magnetic fluxes are given by
\beq
\begin{pmatrix}
  k^1\\
  k^2
\end{pmatrix} =
\begin{pmatrix}
  n_1 - qn_2\\
  n_2
\end{pmatrix},
\eeq
and the signs of $(n_1,n_2)$ (when nonvanishing) must obey
\beq
\sign(n_1) = \sign(\alpha k^1 + \beta k^2),\qquad
\sign(n_2) = \sign(\alpha k^1 + \beta k^2)\sign(\beta - q\alpha),
\eeq
and finally by theorem \ref{thm:3}, $\alpha k^1+\beta k^2\neq0$ cannot
vanish.

We can now make a change of variables:
$u_A=2\log\left|\frac{h_A(z)}{\varsigma_A(z,\bar{z})}\right|$, for
which we can write the system of PDEs as Eq.~\eqref{eq:Taubes_A}:
\beq
\nabla^2u_A = \sum_B \mathcal{A}_{AB}\left(e^{u_B}-1\right)
+ 4\pi\sum_{i=1}^{|n_A|}\delta(z - Z_i^A),
\eeq
where we have defined the real matrix
\beq
\mathcal{A}_{AB} \equiv 2
\begin{pmatrix}
  (e_1^2 + q^2e_2^2)\alpha & qe_2^2(\beta - q\alpha)\\
  \sign(\beta - q\alpha)qe_2^2\alpha & e_2^2|\beta-q\alpha|
\end{pmatrix}.
\eeq
Note that although the diagonal entries are all positive definite, the
matrix $\mathcal{A}$ is not symmetric, but always positive definite.
We can see this from the determinant
\beq
\det\mathcal{A} =
4e_1^2e_2^2\alpha|\beta-q\alpha|>0,
\eeq
which can never switch sign or vanish as $\beta-q\alpha\neq 0$ and
$\alpha>0$.
This is also ($1/4$ times) the determinant of the photon mass-squared
matrix in Eq.~\eqref{eq:detMsq_N=M=2} and the nonvanishing of that
matrix is guaranteed by requiring completely broken gauge symmetry.  

In the special case where the matrix $\mathcal{A}$ is symmetric, the
existence and uniqueness (as well as sharp decay estimates) have been
proved by Yang in Ref.~\cite{Yang:2000in}.
Such a symmetric case corresponds (for this specific example) to:
\beq
\alpha = |\beta-q\alpha|,
\eeq
which has several solutions
\begin{align}
  \beta=0, \quad q=\pm1,\\
  \beta=(q+1)\alpha,
\end{align}
etc.
The origin of this criterion of symmetricity is that
$\sum_aQ_{Aa}e_a^2Q_{Ba}$ of theorem \ref{thm:10} is always a real 
symmetric matrix, but with the addendum of the factor of $z_B$, this
no longer holds true.

\paragraph[Type B2 solution]{Type B2 solution.}

By theorem \ref{thm:4}, the B2 solution has a block diagonal charge
matrix, which for $N=M=2$ means diagonal
\beq
Q =
\begin{pmatrix}
  p & 0\\
  0 & q
\end{pmatrix},\qquad
p,q\in\mathbb{Z}\backslash\{0\},
\label{eq:Q_example_N=M=2_typeB2}
\eeq
and the FI vectors cannot all be parallel and since there are only two
of them $(N=2)$, we take the most generic case of FI vectors (after
the $\SU(2)_R$ symmetry simplification): $\br_1=(\alpha,0,0)$ and
$\br_2=(\beta,\gamma,0)$ with $\alpha>0$, $\beta\in\mathbb{R}$ and
$\gamma\neq0$.
By theorem \ref{thm:6}, the winding flavor is coupled to $\br_1$,
without loss of generality, which in our case means that the winding
flavor is $\tilde{A}=1$.
By corollary \ref{coro:5}, the vacuum of the winding flavor is
\beq
z_1 = \frac{\alpha}{p},
\eeq
and the magnetic flux is given by
\beq
k^1 = \frac{n_1}{p},
\eeq
so the sign $m_1=\sign(\alpha k^1)$ and the sign determining which
field is winding is $\sign(m_1z_1)=\sign(n_1)$, which is unconstrained
(because it is only a single winding flavor in this case).
By theorem \ref{thm:8}, $\alpha k^1\neq0$ cannot vanish and that is
certainly true as we must have $n_1\neq0$.
The non-winding flavors obey the constraint equations by lemma
\ref{lemma:3} and the vortex equation by theorem \ref{thm:5}.
The non-winding of the second flavor, $n_2=0$, implies that $k^2=0$
and hence the string tension reads
\beq
T=T_{\rm BPS} = 2\pi\alpha|k^1|.
\eeq
Finally, by theorem \ref{thm:11}, we have the master equation
\beq
-2\p_{\bar{z}}\p_z\log|\varsigma_1(z,\bar{z})|^2
=|p|e_1^2\alpha\left(\left|\frac{h_1(z)}{\varsigma_1(z,\bar{z})}\right|^2-1\right).
\eeq
Comparing with the master equation \eqref{eq:master_N=M=1}, this case
reduces to the standard Taubes equation and the details are identical
to the case discussed in Sec.~\ref{sec:example_N=M=1}.

It will prove instructive to see explicitly how the constraint
equations are solved in this simple case.
The $\SU(2)_R$ vectors read
\begin{equation}
\bm=(\sign(k^1),0,0), \qquad
\bar\bell=-\frac{1}{\sqrt{2}}(0,\i,\sign(k^1)), \qquad
\bOmega_1=\left(\frac{\alpha}{p},0,0\right), \qquad
\bOmega_2=\left(\frac{\beta}{q},\frac{\gamma}{q},0\right),
\end{equation}
so by the fact that the FI vectors are not parallel and by theorem
\ref{thm:4} this solution type solves the constraint equation
\eqref{eq:constraint_psi2} nontrivially.
As promised by corollary \ref{coro:4}, the winding flavor must have a
vanishing product $\psi_1\tilde\psi_1=0$ as
\beq
\psi_1\tilde\psi_1 = \frac{1}{\sqrt{2}}\bar\bell\cdot\bOmega_1 = 0.
\eeq
As a requirement of theorem \ref{thm:4}, for the non-winding flavor,
the inner product of $\bar\bell$ and $\bOmega_{\hat{A}}$ does not
vanish
\beq
\psi_2\tilde\psi_2 = \frac{1}{\sqrt{2}}\bar\bell\cdot\bOmega_2
= -\frac{\i\gamma}{2q}. 
\eeq
Calculating now the VEV $\langle\psi_2\tilde\psi_2\rangle$, we have
\begin{equation}
\langle\psi_2\tilde\psi_2\rangle = \frac12
\left[\phi_2\tilde\phi_2 - \phi_2^*\tilde\phi_2^*
- \sign(k^1)\left(|\phi_2|^2 - |\tilde\phi_2|^2\right)
\right]
= -\frac12\left[\i\Im(z_2) + \sign(k^1)y_2\right]
= -\frac{\i\gamma}{2q},
\end{equation}
where we have used the useful relations \eqref{eq:useful_vac_rels} and
the vacuum solution for the non-winding flavor
\beq
z_2 = \frac{\beta+\i\gamma}{q}, \qquad y_2=0.
\eeq
This solution remains a solution throughout $\mathbb{R}^2$, because
the magnetic flux $k^2$ is not turned on and the vortex equation for
the winding field decouples due to the diagonality of the charge
matrix \eqref{eq:Q_example_N=M=2_typeB2}.

\paragraph[Type B1 solution]{Type B1 solution.}

By theorem \ref{thm:4}, the B1 solution can have an arbitrary charge
matrix with integer elements and nonvanishing determinant, but only a
single winding flavor.
We take the charge matrix to be
\beq
Q =
\begin{pmatrix}
  1 & q\\
  0 & 1
\end{pmatrix}, \qquad
q\in\mathbb{Z},
\eeq
and the winding flavor to be $\tilde{A}=2$.
Also by theorem \ref{thm:4} there is no restriction on the FI vectors,
so we will take the most general case $\br_1=(\alpha,0,0)$ and 
$\br_2=(\beta,\gamma,0)$, with $\alpha>0$ and $|\br_2|\neq0$.
The vacuum solution is given by
\beq
z_A = (\alpha, \beta - q\alpha),
\eeq
and the magnetic fluxes by
\beq
k^a = (-q n_2, n_2).
\eeq
The string tension reads
\beq
T=T_{\rm BPS}=2\pi\sqrt{(\alpha k^1 + \beta k^2)^2 + (\gamma k^2)^2}
=2\pi\sqrt{(\beta-q\alpha)^2 + \gamma^2}|n_2|,
\eeq
which is linear in the (only) winding number $n_2$, but has a
square-root form of the FI parameters.
By theorem \ref{thm:9}, the vector
\beq
\sum_a\br_a k^a = (\beta - q\alpha, \gamma, 0)n_2,
\eeq
cannot have a vanishing length, i.e.
\beq
\sqrt{(\beta - q\alpha)^2 + \gamma^2}|n_2|\neq 0,
\eeq
and therefore the BPS bound cannot vanish as stated in the theorem.
The $\bm$ vector is thus
\beq
\bm = \frac{(\beta-q\alpha,\gamma,0)}{\sqrt{(\beta-q\alpha)^2+\gamma^2}}\sign(n_2).
\eeq
Finally, by theorem \ref{thm:12}, the master equations read
\begin{align}
  -2\p_{\bar{z}}\p_z\log|\varsigma_1(z,\bar{z})|^2
  &=\sign(\beta-q\alpha)qe_2^2\sqrt{(\beta-q\alpha)^2+\gamma^2}
  \left(\left|\frac{h_2(z)}{\varsigma_2(z,\bar{z})}\right|^2-1\right)
  \non
  &\phantom{=\ }
  +\frac{e_1^2+q^2e_2^2}{2}\Bigg[
    \left(\alpha +
    \frac{\alpha|\beta-q\alpha|}{\sqrt{(\beta-q\alpha)^2+\gamma^2}}\right)\left(\frac{1}{|\varsigma_1(z,\bar{z})|^2}-1\right) \non&\phantom{=\ }
    -\left(\alpha -
    \frac{\alpha|\beta-q\alpha|}{\sqrt{(\beta-q\alpha)^2+\gamma^2}}\right)\left(|\varsigma_1(z,\bar{z})|^2-1\right)\Bigg],\\
  -2\p_{\bar{z}}\p_z\log|\varsigma_2(z,\bar{z})|^2
  &=e_2^2\sqrt{(\beta-q\alpha)^2+\gamma^2}
  \left(\left|\frac{h_2(z)}{\varsigma_2(z,\bar{z})}\right|^2-1\right)
  \non
  &\phantom{=\ }
  +\frac{q e_2^2}{2}\Bigg[
    \left(\sign(\beta-q\alpha)\alpha +
    \frac{\alpha(\beta-q\alpha)}{\sqrt{(\beta-q\alpha)^2+\gamma^2}}\right)\left(\frac{1}{|\varsigma_1(z,\bar{z})|^2}-1\right) \non&\phantom{=\ }
    -\left(\sign(\beta-q\alpha)\alpha -
    \frac{\alpha(\beta-q\alpha)}{\sqrt{(\beta-q\alpha)^2+\gamma^2}}\right)\left(|\varsigma_1(z,\bar{z})|^2-1\right)\Bigg],
\end{align}
where $\beta-q\alpha\neq0$ cannot vanish as that would imply unbroken
gauge symmetry.

We can now make a change of variables:
$u_1=-2\log|\varsigma_1(z,\bar{z})|$,
$u_2=2\log\left|\frac{h_2(z)}{\varsigma_2(z,\bar{z})}\right|$, for
which we can write the system of PDEs as Eq.~\eqref{eq:Taubes_B1}:
\begin{align}
  \nabla^2u_A = \mathcal{A}_{A}(e^{u_2}-1)
  +\sum_{B\neq2}\mathcal{B}_{AB}(e^{u_B}-1)
  -\sum_{B\neq2}\mathcal{C}_{AB}(e^{-n_B}-1)
  + 4\pi\delta^{A2}\sum_{i=1}^{|n_2|}\delta(z-Z_i),
  \label{eq:Taubes_N=M=2_typeB1}
\end{align}
where we have defined
\begin{align}
  \mathcal{A}_A &\equiv 2
  \begin{pmatrix}
    \sign(\beta-q\alpha)qe_2^2\sqrt{(\beta-q\alpha)^2+\gamma^2}\\
    e_2^2\sqrt{(\beta-q\alpha)^2+\gamma^2}
  \end{pmatrix},\\
  \mathcal{B}_{AB} &\equiv
  \begin{pmatrix}
    (e_1^2+q^2e_2^2)
    \left(\alpha +
    \frac{\alpha|\beta-q\alpha|}{\sqrt{(\beta-q\alpha)^2+\gamma^2}}\right)\\
    q e_2^2
    \left(\sign(\beta-q\alpha)\alpha +
    \frac{\alpha(\beta-q\alpha)}{\sqrt{(\beta-q\alpha)^2+\gamma^2}}\right)
  \end{pmatrix},\\
  \mathcal{C}_{AB} &\equiv
  \begin{pmatrix}
    (e_1^2+q^2e_2^2)
    \left(\alpha -
    \frac{\alpha|\beta-q\alpha|}{\sqrt{(\beta-q\alpha)^2+\gamma^2}}\right)\\
    q e_2^2
    \left(\sign(\beta-q\alpha)\alpha -
    \frac{\alpha(\beta-q\alpha)}{\sqrt{(\beta-q\alpha)^2+\gamma^2}}\right)
  \end{pmatrix},   
\end{align}
with $\mathcal{A}$ an $N$-vector and $\mathcal{B}$ and
$\mathcal{C}$ are $N$-by-$(N-1)$ matrices (and $N=2$). 
We do not know of any existence or uniqueness results for the equation
\eqref{eq:Taubes_N=M=2_typeB1}.

\subsubsection{\texorpdfstring{$N=M=3$}{N=M=3}}

We now turn to the case of three gauge groups with three flavors
($N=M=3$).

\paragraph[Type A solution]{Type A solution.}

By theorem \ref{thm:1}, the FI vectors are parallel and hence read
$\br_1=(\alpha,0,0)$, $\br_2=(\beta,0,0)$ and $\br_3=(\delta,0,0)$ with
$\alpha>0$, and $\beta,\delta\neq 0$ and in turn $\bm=(\pm1,0,0)$ (by
theorem \ref{thm:1}).
The vacuum solution is given by $z_A=\sum_ar_a^1Q^{-1}_{aA}$
(corollary \ref{coro:2}) and the string tension reads 
\beq
T_{\rm BPS} = 2\pi|\alpha k^1 + \beta k^2 + \delta k^3|>0.
\eeq
We will now select an explicit example for the charge matrix with
nonvanishing determinant
\begin{align}
Q =
\begin{pmatrix}
  1 & p & q\\ 
  0 & 1 & r\\
  0 & 0 & 1
\end{pmatrix}, \qquad
p,q,r\in\mathbb{Z},
\end{align}
for which the vacuum solution is given by
\beq
z_A = (\alpha, \beta - p\alpha, (pr - q)\alpha - r\beta + \delta),
\eeq
and the master equations by theorem \ref{thm:10} read
\begin{align}
-2\p_{\bar{z}}\p_z\log|\varsigma_1(z,\bar{z})|^2
&= (e_1^2+p^2e_2^2+q^2e_3^2)\alpha\left(\left|\frac{h_1(z)}{\varsigma_1(z,\bar{z})}\right|^2-1\right)\non
&\phantom{=\ }
+(pe_2^2+qre_3^2)(\beta-p\alpha)\left(\left|\frac{h_2(z)}{\varsigma_2(z,\bar{z})}\right|^2-1\right)\non
&\phantom{=\ }
+qe_3^2((pr-q)\alpha-r\beta+\delta)\left(\left|\frac{h_3(z)}{\varsigma_3(z,\bar{z})}\right|^2-1\right),\\
-2\p_{\bar{z}}\p_z\log|\varsigma_2(z,\bar{z})|^2
&= \sign(\beta-p\alpha)(pe_2^2+qre_3^2)\alpha\left(\left|\frac{h_1(z)}{\varsigma_1(z,\bar{z})}\right|^2-1\right)\non
&\phantom{=\ }
+(e_2^2+r^2e_3^2)|\beta-p\alpha|\left(\left|\frac{h_2(z)}{\varsigma_2(z,\bar{z})}\right|^2-1\right)\non
&\phantom{=\ }
+\sign(\beta-p\alpha)re_3^2((pr-q)\alpha-r\beta+\delta)\left(\left|\frac{h_3(z)}{\varsigma_3(z,\bar{z})}\right|^2-1\right),\\
-2\p_{\bar{z}}\p_z\log|\varsigma_3(z,\bar{z})|^2
&= \sign((pr-q)\alpha-r\beta+\delta)qe_3^2\alpha\left(\left|\frac{h_1(z)}{\varsigma_1(z,\bar{z})}\right|^2-1\right)\non
&\phantom{=\ }
+\sign((pr-q)\alpha-r\beta+\delta)re_3^2(\beta-p\alpha)\left(\left|\frac{h_2(z)}{\varsigma_2(z,\bar{z})}\right|^2-1\right)\non
&\phantom{=\ }
+e_3^2|(pr-q)\alpha-r\beta+\delta|\left(\left|\frac{h_3(z)}{\varsigma_3(z,\bar{z})}\right|^2-1\right).
\end{align}
Note that $\beta-p\alpha\neq0$ and $(pr-q)\alpha-r\beta+\delta\neq0$
cannot vanish as that would make the theory touch the Coulomb branch
on the second and third flavors, respectively, or in other words, the
gauge symmetry would be unbroken.
  
By theorem \ref{thm:2}, the magnetic fluxes are given by
\beq
\begin{pmatrix}
  k^1\\
  k^2\\
  k^3
\end{pmatrix} =
\begin{pmatrix}
  n_1 - pn_2 + (pr-q)n_3\\
  n_2 - rn_3\\
  n_3
\end{pmatrix},
\eeq
and the signs of $(n_1,n_2,n_3)$ when nonvanishing must obey
\begin{align}
\sign(n_1) &= \sign(\alpha k^1 + \beta k^2 + \delta k^3),\\
\sign(n_2) &= \sign(\alpha k^1 + \beta k^2 + \delta k^3)
\sign(\beta - p\alpha),\\
\sign(n_3) &= \sign(\alpha k^1 + \beta k^2 + \delta k^3)
\sign((pr-q)\alpha - r\beta + \delta),
\end{align}
and finally by theorem \ref{thm:3},
$\alpha k^1+\beta k^2+\delta k^3\neq0$ cannot vanish.

We can now make a change of variables:
$u_A=2\log\left|\frac{h_A(z)}{\varsigma_A(z,\bar{z})}\right|$, for
which we can write the system of PDEs as Eq.~\eqref{eq:Taubes_A}:
\beq
\nabla^2u_A = \sum_B \mathcal{A}_{AB}\left(e^{u_B}-1\right)
+ 4\pi\sum_{i=1}^{|n_A|}\delta(z - Z_i^A),
\eeq
where we have defined the real matrix
\beq
\mathcal{A}_{AB} \equiv 2
\begin{pmatrix}
  (e_1^2+p^2e_2^2+q^2e_3^2)\alpha &
  (pe_2^2+qre_3^2)(\beta-p\alpha) & 
  qe_3^2((pr-q)\alpha-r\beta+\delta)\\
  \mathcal{S}_2(pe_2^2+qre_3^2)\alpha &
  (e_2^2+r^2e_3^2)|\beta-p\alpha| &
  \mathcal{S}_2re_3^2((pr-q)\alpha-r\beta+\delta)\\
  \mathcal{S}_3qe_3^2\alpha &
  \mathcal{S}_3re_3^2(\beta-p\alpha) &
  e_3^2|(pr-q)\alpha-r\beta+\delta|
\end{pmatrix},
\eeq
with
\beq
\mathcal{S}_2 \equiv \sign(\beta-p\alpha), \qquad
\mathcal{S}_3 \equiv \sign((pr-q)\alpha-r\beta+\delta).
\eeq
Note that although the diagonal entries are all positive definite, the
matrix $\mathcal{A}$ is not symmetric, but always positive definite.
We can see this from the determinant
\beq
\det\mathcal{A} =
8e_1^2e_2^2e_3^2\alpha\left|(\beta-p\alpha)((pr-q)\alpha-r\beta+\delta)\right|>0,
\eeq
which can never switch sign or vanish as $\beta-p\alpha\neq 0$,
$(pr-q)\alpha-r\beta+\delta\neq0$ and $\alpha>0$.
This is also ($1/8$ times) the determinant of the mass-squared matrix
\eqref{eq:detMsqN=M=3} for the photons, which must be positive
definite in order to ensure broken gauge symmetry.

Similarly to the $N=M=2$ example, in the special case where the matrix
$\mathcal{A}$ is symmetric, the existence and uniqueness (as well as
sharp decay estimates) have been proven by Yang in
Ref.~\cite{Yang:2000in} (same system).
Such a symmetric case corresponds (for this specific example) to:
\beq
\alpha = |\beta-p\alpha|, \qquad
\alpha = |(pr - q)\alpha - r\beta + \delta|.
\eeq

\paragraph[Type B2 solution]{Type B2 solution.}

By theorem \ref{thm:4}, the B2 solution has a block diagonal charge
matrix, which in this example we shall choose in the form
\beq
Q =
\begin{pmatrix}
  1 & q & 0\\
  0 & 1 & 0\\
  0 & 0 & 1
\end{pmatrix},\qquad
q\in\mathbb{Z}\backslash\{0\},
\label{eq:Q_example_N=M=3_typeB2}
\eeq
and the FI vectors cannot all be parallel by theorem \ref{thm:4}, but
by theorem \ref{thm:6}, the first two FI vectors (due to the winding
block of the above matrix being 2-by-2) must be parallel.
We thus set the FI vectors to be : $\br_1=(\alpha,0,0)$,
$\br_2=(\beta,0,0)$ and $\br_3=(\delta,\kappa,\eta)$ with $\alpha>0$,
$\beta\neq0$ and $|\br_3|\neq0$. 
The winding flavors are $\tilde{A}=1,2$.
By corollary \ref{coro:5}, the vacuum of the winding flavors is
\beq
z_{1,2} = (\alpha,\beta-q\alpha),
\eeq
and the magnetic flux is given by
\beq
k^{1,2} = (n_1-qn_2,n_2),
\eeq
so the sign
\beq
m_1=\sign(\alpha k^1 + \beta k^2).
\eeq
By theorem \ref{thm:8}, $\alpha k^1+\beta k^2\neq0$ cannot vanish.
The non-winding flavors obey the constraint equations by lemma
\ref{lemma:3} and the vortex equations by theorem \ref{thm:5}.
The non-winding of the third flavor, $n_3=0$, implies that $k^3=0$
and hence the string tension reads
\beq
T=T_{\rm BPS} = 2\pi|\alpha k^1 + \beta k^2|.
\eeq
Finally, by theorem \ref{thm:11}, we have the master equations
\begin{align}
-2\p_{\bar{z}}\p_z\log|\varsigma_1(z,\bar{z})|^2
&=(e_1^2+q^2e_2^2)\alpha\left(\left|\frac{h_1(z)}{\varsigma_1(z,\bar{z})}\right|^2-1\right)
+qe_2^2(\beta-q\alpha)\left(\left|\frac{h_2(z)}{\varsigma_2(z,\bar{z})}\right|^2-1\right),\\
-2\p_{\bar{z}}\p_z\log|\varsigma_2(z,\bar{z})|^2
&=\sign(\beta-q\alpha)qe_2^2\alpha\left(\left|\frac{h_1(z)}{\varsigma_1(z,\bar{z})}\right|^2-1\right)
+e_2^2|\beta-q\alpha|\left(\left|\frac{h_2(z)}{\varsigma_2(z,\bar{z})}\right|^2-1\right).
\end{align}
Comparing with the master equations \eqref{eq:master_N=M=2_1} and
\eqref{eq:master_N=M=2_2}, this case reduces to the equations found in
Sec.~\ref{sec:example_N=M=2_type}.

\paragraph[Type B1 solution]{Type B1 solution.}

By theorem \ref{thm:4}, the B1 solution can have an arbitrary charge
matrix with integer elements and nonvanishing determinant, but only a
single winding flavor.
We take the charge matrix to be
\beq
Q =
\begin{pmatrix}
  1 & p & q\\
  0 & 1 & r\\
  0 & 0 & 1
\end{pmatrix}, \qquad
p,q,r\in\mathbb{Z}\backslash\{0\},
\eeq
and the winding flavor to be $\tilde{A}=3$.
Also by theorem \ref{thm:4}, there is no restriction on the FI vectors,
so we will take the most general case $\br_1=(\alpha,0,0)$,
$\br_2=(\beta,\gamma,0)$ and $\br_3=(\delta,\kappa,\eta)$, with
$\alpha>0$, $|\br_2|\neq0$ and $|\br_3|\neq0$.
The vacuum solution is given by
\beq
z_A =
\begin{pmatrix}
  \alpha\\
  \beta - p\alpha + \i\gamma\\
  (pr-q)\alpha - r(\beta+\i\gamma) + \delta + \i\kappa
\end{pmatrix}^{\rm T},\qquad
y_A =
\begin{pmatrix}
  0\\
  0\\
  \eta
\end{pmatrix}^{\rm T},
\eeq
and the magnetic fluxes by
\beq
k^a =
\begin{pmatrix}
  (pr-q)n_3\\
  -r n_3\\
  n_3
\end{pmatrix}.
\eeq
The string tension reads
\begin{align}
T=T_{\rm BPS}&=2\pi\sqrt{(\alpha k^1 + \beta k^2 + \delta k^3)^2 +
  (\gamma k^2 + \kappa k^3)^2 + (\eta k^3)^2} \non
&=2\pi\sqrt{((pr-q)\alpha - r\beta + \delta)^2 +
  (\kappa - r\gamma)^2 + \eta^2}|n_3|,
\end{align}
which is linear in the (only) winding number $n_3$, but has a
square-root form of the FI parameters.
By theorem \ref{thm:9}, the vector
\beq
\sum_a\br_a k^a =
\begin{pmatrix}
  (pr-q)\alpha - r\beta + \delta\\
  \kappa - r\gamma\\
  \eta
\end{pmatrix} n_3
\eeq
cannot have a vanishing length, i.e.
\beq
\sqrt{((pr-q)\alpha - r\beta + \delta)^2
  + (\kappa - r\gamma)^2 + \eta^2}|n_3| \neq 0,
\eeq
and therefore the BPS bound cannot vanish as stated in the theorem.
The $\bm$ vector reads
\beq
\bm =
\frac{1}{\sqrt{((pr-q)\alpha - r\beta + \delta)^2
  + (\kappa - r\gamma)^2 + \eta^2}}
\begin{pmatrix}
  (pr-q)\alpha - r\beta + \delta\\
  \kappa - r\gamma\\
  \eta
\end{pmatrix}
\sign(n_3).
\eeq
Finally, by theorem \ref{thm:12}, the master equations read
\begin{align}
  &-2\p_{\bar{z}}\p_z\log|\varsigma_1(z,\bar{z})|^2
  =\sign(n_3)qe_3^2
  \sqrt{((pr-q)\alpha-r\beta+\delta)^2+(\kappa-r\gamma)^2+\eta^2}
  \left(\left|\frac{h_3(z)}{\varsigma_3(z,\bar{z})}\right|^2-1\right)
  \non
  &
  +\frac{e_1^2+p^2e_2^2+q^2e_3^2}{2}\Bigg[
    \left(\alpha +
    \frac{\sign(n_3)\alpha((pr-q)\alpha-r\beta+\delta)}{\sqrt{((pr-q)\alpha-r\beta+\delta)^2+(\kappa-r\gamma)^2+\eta^2}}\right)
    \left(\frac{1}{|\varsigma_1(z,\bar{z})|^2}-1\right) \non&
    -\left(\alpha -
    \frac{\sign(n_3)\alpha((pr-q)\alpha-r\beta+\delta)}{\sqrt{((pr-q)\alpha-r\beta+\delta)^2+(\kappa-r\gamma)^2+\eta^2}}\right)
    \left(|\varsigma_1(z,\bar{z})|^2-1\right)\Bigg]\non&
  +\frac{pe_2^2+q r e_3^2}{2}\Bigg[
    \Bigg(\sqrt{(\beta-p\alpha)^2+\gamma^2}\non&
    +\sign(n_3)
    \frac{(\beta-p\alpha)((pr-q)\alpha-r\beta+\delta)+\gamma(\kappa-r\gamma)}{\sqrt{((pr-q)\alpha-r\beta+\delta)^2+(\kappa-r\gamma)^2+\eta^2}}\Bigg)
    \left(\frac{1}{|\varsigma_2(z,\bar{z})|^2}-1\right) \non&
    -\left(\sqrt{(\beta-p\alpha)^2+\gamma^2} - \sign(n_3)
    \frac{(\beta-p\alpha)((pr-q)\alpha-r\beta+\delta)+\gamma(\kappa-r\gamma)}{\sqrt{((pr-q)\alpha-r\beta+\delta)^2+(\kappa-r\gamma)^2+\eta^2}}\right)
    \left(|\varsigma_2(z,\bar{z})|^2-1\right)\Bigg],
\end{align}
\begin{align}
  &-2\p_{\bar{z}}\p_z\log|\varsigma_2(z,\bar{z})|^2
  =\sign(n_3)re_3^2
  \sqrt{((pr-q)\alpha-r\beta+\delta)^2+(\kappa-r\gamma)^2+\eta^2}
  \left(\left|\frac{h_3(z)}{\varsigma_3(z,\bar{z})}\right|^2-1\right)
  \non
  &
  +\frac{pe_2^2+qre_3^2}{2}\Bigg[
    \left(\alpha +
    \frac{\sign(n_3)\alpha((pr-q)\alpha-r\beta+\delta)}{\sqrt{((pr-q)\alpha-r\beta+\delta)^2+(\kappa-r\gamma)^2+\eta^2}}\right)
    \left(\frac{1}{|\varsigma_1(z,\bar{z})|^2}-1\right) \non&
    -\left(\alpha -
    \frac{\sign(n_3)\alpha((pr-q)\alpha-r\beta+\delta)}{\sqrt{((pr-q)\alpha-r\beta+\delta)^2+(\kappa-r\gamma)^2+\eta^2}}\right)
    \left(|\varsigma_1(z,\bar{z})|^2-1\right)\Bigg]\non&
  +\frac{e_2^2+r^2e_3^2}{2}\Bigg[
    \Bigg(\sqrt{(\beta-p\alpha)^2+\gamma^2}\non&
    +\sign(n_3)
    \frac{(\beta-p\alpha)((pr-q)\alpha-r\beta+\delta)+\gamma(\kappa-r\gamma)}{\sqrt{((pr-q)\alpha-r\beta+\delta)^2+(\kappa-r\gamma)^2+\eta^2}}\Bigg)
    \left(\frac{1}{|\varsigma_2(z,\bar{z})|^2}-1\right) \non&
    -\left(\sqrt{(\beta-p\alpha)^2+\gamma^2} - \sign(n_3)
    \frac{(\beta-p\alpha)((pr-q)\alpha-r\beta+\delta)+\gamma(\kappa-r\gamma)}{\sqrt{((pr-q)\alpha-r\beta+\delta)^2+(\kappa-r\gamma)^2+\eta^2}}\right)
    \left(|\varsigma_2(z,\bar{z})|^2-1\right)\Bigg],
\end{align}
\begin{align}
  &-2\p_{\bar{z}}\p_z\log|\varsigma_3(z,\bar{z})|^2
  =e_3^2
  \sqrt{((pr-q)\alpha-r\beta+\delta)^2+(\kappa-r\gamma)^2+\eta^2}
  \left(\left|\frac{h_3(z)}{\varsigma_3(z,\bar{z})}\right|^2-1\right)
  \non
  &
  +\frac{qe_3^2}{2}\Bigg[
    \left(\sign(n_3)\alpha +
    \frac{\alpha((pr-q)\alpha-r\beta+\delta)}{\sqrt{((pr-q)\alpha-r\beta+\delta)^2+(\kappa-r\gamma)^2+\eta^2}}\right)
    \left(\frac{1}{|\varsigma_1(z,\bar{z})|^2}-1\right) \non&
    -\left(\sign(n_3)\alpha -
    \frac{\alpha((pr-q)\alpha-r\beta+\delta)}{\sqrt{((pr-q)\alpha-r\beta+\delta)^2+(\kappa-r\gamma)^2+\eta^2}}\right)
    \left(|\varsigma_1(z,\bar{z})|^2-1\right)\Bigg]\non&
  +\frac{re_3^2}{2}\Bigg[
    \Bigg(\sign(n_3)\sqrt{(\beta-p\alpha)^2+\gamma^2}\non&
    +
    \frac{(\beta-p\alpha)((pr-q)\alpha-r\beta+\delta)+\gamma(\kappa-r\gamma)}{\sqrt{((pr-q)\alpha-r\beta+\delta)^2+(\kappa-r\gamma)^2+\eta^2}}\Bigg)
    \left(\frac{1}{|\varsigma_2(z,\bar{z})|^2}-1\right) \non&
    -\left(\sign(n_3)\sqrt{(\beta-p\alpha)^2+\gamma^2} - 
    \frac{(\beta-p\alpha)((pr-q)\alpha-r\beta+\delta)+\gamma(\kappa-r\gamma)}{\sqrt{((pr-q)\alpha-r\beta+\delta)^2+(\kappa-r\gamma)^2+\eta^2}}\right)
    \left(|\varsigma_2(z,\bar{z})|^2-1\right)\Bigg],
\end{align}
where $(\beta-p\alpha)^2+\gamma^2\neq0$ and
$((pr-q)\alpha-r\beta+\delta)^2+(\kappa-r\gamma)^2\neq0$ cannot vanish
as either would imply unbroken gauge symmetry.

We can now make a change of variables:
$u_1=-2\log|\varsigma_1(z,\bar{z})|$,
$u_2=-2\log|\varsigma_2(z,\bar{z})|$,
$u_3=2\log\left|\frac{h_3(z)}{\varsigma_3(z,\bar{z})}\right|$, for
which we can write the system of PDEs as Eq.~\eqref{eq:Taubes_B1}:
\begin{align}
  \nabla^2u_A = \mathcal{A}_{A}(e^{u_3}-1)
  +\sum_{B\neq3}\mathcal{B}_{AB}(e^{u_B}-1)
  -\sum_{B\neq3}\mathcal{C}_{AB}(e^{-n_B}-1)
  + 4\pi\delta^{A3}\sum_{i=1}^{|n_3|}\delta(z-Z_i),
  \label{eq:Taubes_N=M=3_typeB1}
\end{align}
where we have defined
\begin{align}
  \mathcal{A}_A &\equiv 2
  \begin{pmatrix}
    \sign(n_3)qe_3^2\\
    \sign(n_3)re_3^2\\
    e_3^2
  \end{pmatrix}
  \sqrt{((pr-q)\alpha-r\beta+\delta)^2+(\kappa-r\gamma)^2+\eta^2},
\end{align}
\begin{align}
  \mathcal{B}_{A1} &\equiv
  \begin{pmatrix}
    e_1^2+p^2e_2^2+q^2e_3^2\\
    pe_2^2+qre_3^2\\
    \sign(n_3)qe_3^2
  \end{pmatrix}
  \left(\alpha + \frac{\sign(n_3)\alpha((pr-q)\alpha-r\beta+\delta)}{\sqrt{((pr-q)\alpha-r\beta+\delta)^2+(\kappa-r\gamma)^2+\eta^2}}\right),\\
  \mathcal{B}_{A2} &\equiv \non&
  \begin{pmatrix}
    pe_2^2+q r e_3^2\\
    e_2^2+r^2e_3^2\\
    \sign(n_3)re_3^2
  \end{pmatrix}
  \left(\sqrt{(\beta-p\alpha)^2+\gamma^2}
    +\sign(n_3)
    \frac{(\beta-p\alpha)((pr-q)\alpha-r\beta+\delta)+\gamma(\kappa-r\gamma)}{\sqrt{((pr-q)\alpha-r\beta+\delta)^2+(\kappa-r\gamma)^2+\eta^2}}\right),
\end{align}
\begin{align}
  \mathcal{C}_{A1} &\equiv
  -\begin{pmatrix}
    e_1^2+p^2e_2^2+q^2e_3^2\\
    pe_2^2+qre_3^2\\
    \sign(n_3)qe_3^2
  \end{pmatrix}
  \left(\alpha - \frac{\sign(n_3)\alpha((pr-q)\alpha-r\beta+\delta)}{\sqrt{((pr-q)\alpha-r\beta+\delta)^2+(\kappa-r\gamma)^2+\eta^2}}\right),\\
  \mathcal{C}_{A2} &\equiv\non&
  \begin{pmatrix}
    pe_2^2+q r e_3^2\\
    e_2^2+r^2e_3^2\\
    \sign(n_3)re_3^2
  \end{pmatrix}
  \left(\sqrt{(\beta-p\alpha)^2+\gamma^2}
    -\sign(n_3)
    \frac{(\beta-p\alpha)((pr-q)\alpha-r\beta+\delta)+\gamma(\kappa-r\gamma)}{\sqrt{((pr-q)\alpha-r\beta+\delta)^2+(\kappa-r\gamma)^2+\eta^2}}\right),
\end{align}
with $\mathcal{A}$ an $N$-vector and $\mathcal{B}$ and
$\mathcal{C}$ are $N$-by-$(N-1)$ matrices (and $N=3$). 
We do not know of any existence or uniqueness results for the equation
\eqref{eq:Taubes_N=M=3_typeB1}.

\subsection*{Acknowledgments}

S.~B.~G.~thanks Xiaosen Han, Keisuke Ohashi, Calum Ross and Yisong
Yang for discussions and correspondence. 
S.~B.~G.~thanks the Outstanding Talent Program of Henan University for
partial support.
The work of S.~B.~G.~is supported by the National Natural Science
Foundation of China (Grant No.~11675223 and 12071111).
The work of M.~E.~is supported in part by JSPS Grant-in-Aid for
Scientific Research (KAKENHI Grant No. JP19K03839, and also by
MEXT KAKENHI Grant-in-Aid for Scientific Research on Innovative Areas,
Discrete Geometric Analysis for Materials Design, No.~JP17H06462 from
the MEXT of Japan. 
The work of M.N.~is supported in part by JSPS Grant-in-Aid for
Scientific Research (KAKENHI Grant No.~18H01217).


\begin{thebibliography}{99}

\bibitem{Polchinski:1995mt}
J.~Polchinski,
``Dirichlet Branes and Ramond-Ramond charges,''
\href{http://dx.doi.org/10.1103/PhysRevLett.75.4724}{Phys. Rev. Lett. \textbf{75}, 4724-4727 (1995)}
[\href{http://www.arxiv.org/abs/arXiv:hep-th/9510017}{arXiv:hep-th/9510017 [hep-th]}].

\bibitem{Witten:1995im}
E.~Witten,
``Bound states of strings and p-branes,''
\href{http://dx.doi.org/10.1016/0550-3213(95)00610-9}{Nucl. Phys. B \textbf{460}, 335-350 (1996)}
[\href{http://www.arxiv.org/abs/arXiv:hep-th/9510135}{arXiv:hep-th/9510135 [hep-th]}].

\bibitem{Copeland:2003bj}
E.~J.~Copeland, R.~C.~Myers and J.~Polchinski,
``Cosmic F and D strings,''
\href{http://dx.doi.org/10.1088/1126-6708/2004/06/013}{JHEP \textbf{06}, 013 (2004)}
[\href{http://www.arxiv.org/abs/arXiv:hep-th/0312067}{arXiv:hep-th/0312067 [hep-th]}].

\bibitem{Polchinski:2004ia}
J.~Polchinski,
``Introduction to cosmic F- and D-strings,''
[\href{http://www.arxiv.org/abs/arXiv:hep-th/0412244}{arXiv:hep-th/0412244 [hep-th]}].

\bibitem{Polchinsky:1998}
J.~Polchinsky,
``String Theory -- Volume II: Superstring Theory and Beyond,''
Cambridge University Press (1998).

\bibitem{Saffin:2005cs}
P.~M.~Saffin,
``A Practical model for cosmic (p,q) superstrings,''
\href{http://dx.doi.org/10.1088/1126-6708/2005/09/011}{JHEP \textbf{09}, 011 (2005)}
[\href{http://www.arxiv.org/abs/arXiv:hep-th/0506138}{arXiv:hep-th/0506138 [hep-th]}].

\bibitem{Jackson:2006qc}
M.~G.~Jackson,
``A Note on Cosmic (p,q,r) Strings,''
\href{http://dx.doi.org/10.1103/PhysRevD.75.087301}{Phys. Rev. D \textbf{75}, 087301 (2007)}
[\href{http://www.arxiv.org/abs/arXiv:hep-th/0610059}{arXiv:hep-th/0610059 [hep-th]}].

\bibitem{Schroers:1996zy}
B.~J.~Schroers,
``The Spectrum of Bogomol'nyi solitons in gauged linear sigma models,''
\href{http://dx.doi.org/10.1016/0550-3213(96)00348-3}{Nucl. Phys. B \textbf{475}, 440-468 (1996)}
[\href{http://www.arxiv.org/abs/hep-th/9603101}{arXiv:hep-th/9603101 [hep-th]}].

\bibitem{Vachaspati:1991dz}
T.~Vachaspati and A.~Achucarro,
``Semilocal cosmic strings,''
\href{http://dx.doi.org/10.1103/PhysRevD.44.3067}{Phys. Rev. D \textbf{44}, 3067-3071 (1991)}.

\bibitem{Eto:2005sw}
M.~Eto, Y.~Isozumi, M.~Nitta and K.~Ohashi,
``1/2, 1/4 and 1/8 BPS equations in SUSY Yang-Mills-Higgs systems: Field theoretical brane configurations,''
\href{http://dx.doi.org/10.1016/j.nuclphysb.2006.06.026}{Nucl. Phys. B \textbf{752}, 140-172 (2006)}
[\href{http://www.arxiv.org/abs/arXiv:hep-th/0506257}{arXiv:hep-th/0506257 [hep-th]}].

\bibitem{Isozumi:2004vg}
Y.~Isozumi, M.~Nitta, K.~Ohashi and N.~Sakai,
``All exact solutions of a 1/4 Bogomol'nyi-Prasad-Sommerfield equation,''
\href{http://dx.doi.org/10.1103/PhysRevD.71.065018}{Phys. Rev. D \textbf{71}, 065018 (2005)}
[\href{http://www.arxiv.org/abs/arXiv:hep-th/0405129}{arXiv:hep-th/0405129 [hep-th]}].

\bibitem{Eto:2005yh}
M.~Eto, Y.~Isozumi, M.~Nitta, K.~Ohashi and N.~Sakai,
``Moduli space of non-Abelian vortices,''
\href{http://dx.doi.org/10.1103/PhysRevLett.96.161601}{Phys. Rev. Lett. \textbf{96}, 161601 (2006)}
[\href{http://www.arxiv.org/abs/arXiv:hep-th/0511088}{arXiv:hep-th/0511088 [hep-th]}].

\bibitem{Eto:2006pg}
M.~Eto, Y.~Isozumi, M.~Nitta, K.~Ohashi and N.~Sakai,
``Solitons in the Higgs phase: The Moduli matrix approach,''
\href{http://dx.doi.org/10.1088/0305-4470/39/26/R01}{J. Phys. A \textbf{39}, R315-R392 (2006)}
[\href{http://www.arxiv.org/abs/arXiv:hep-th/0602170}{arXiv:hep-th/0602170 [hep-th]}].

\bibitem{Yang:2000in}
Y.~Yang,
``On a system of nonlinear elliptic equations arising in theoretical physics,''
\href{http://dx.doi.org/10.1006/jfan.1999.3492}{J. Funct. Anal. \textbf{170}, 1-36 (2000)}.

\bibitem{Yang:2004}
Y.~Yang,
``Solitons in field theory and nonlinear analysis,''
Springer Monographs in Mathematics, Springer (2001). 

\bibitem{Taubes:1979tm}
C.~H.~Taubes,
``Arbitrary N: Vortex Solutions to the First Order Landau-Ginzburg Equations,''
\href{http://dx.doi.org/10.1007/BF01197552}{Commun. Math. Phys. \textbf{72}, 277-292 (1980)}.


  
\end{thebibliography}
\end{document}